\documentclass[aps, prd, 10pt, notitlepage, superscriptaddress, nofootinbib,numbers,showpacs,showkeys,onecolumn,preprintnumbers]{revtex4-2}
\usepackage{graphicx,color}
\usepackage{amsmath}
\usepackage{amssymb}
\usepackage{amsfonts}
\usepackage{bm}
\usepackage{cancel}
\usepackage{slashed}
\usepackage{hyperref}

\usepackage{tikz}
\usepackage{subcaption}

\newcommand{\e}{\mathrm{e}}
\newcommand{\rt}{r_\mathrm{throat}}
\newcommand{\rp}{r_\mathrm{ph}}
\newcommand{\ri}{r_\mathrm{inner}}
\newcommand{\rhor}{r_\mathrm{hor}}
\newcommand{\arcsinh}{\mathrm{arcsinh}}

\allowdisplaybreaks[4]
\begin{document}

\preprint{KEK-TH-2861, KEK-Cosmo-0429}

\title{Horizon singularity, energy conditions and shadows in time-dependent and spherically symmetric spacetime}

\author{Shin'ichi~Nojiri}
\email{nojiri@nagoya-u.jp}
\affiliation{ KEK Theory Center, Institute of Particle and Nuclear Studies,
High Energy Accelerator Research Organization (KEK), Oho 1-1, Tsukuba, Ibaraki 305-0801, Japan}
\affiliation{ Kobayashi-Maskawa Institute for the Origin of Particles and the Universe, Nagoya University, Nagoya 464-8602, Japan}

\author{Sergei~D.~Odintsov}
\email{odintsov@ice.csic.es} 
\affiliation{Institute of Space Sciences (ICE-CSIC) C. Can Magrans s/n, 08193 Barcelona, Spain}
\affiliation{ Instituci\'o Catalana de Recerca i Estudis Avan\c{c}ats (ICREA),
Passeig Luis Companys, 23, 08010 Barcelona, Spain}

\author{Diego~S\'aez-Chill\'on~G\'omez}
\email{diego.saez@uva.es}
\affiliation{ Department of Theoretical, Atomic and Optical
Physics, and Laboratory for Disruptive Interdisciplinary Science (LaDIS), Campus Miguel Delibes, \\ 
University of Valladolid UVA, Paseo Bel\'en, 7, 47011
Valladolid, Spain}
\affiliation{ Department of Physics, Universidade Federal do Cear\'a (UFC), Campus do Pici, Fortaleza - CE, C.P. 6030, 60455-760 - Brazil}

\author{Álvaro Salazar Cuadros}
\email{a.salazar@uva.es}
\affiliation{ Department of Theoretical, Atomic and Optical
Physics, and Laboratory for Disruptive Interdisciplinary Science (LaDIS), Campus Miguel Delibes, \\ 
University of Valladolid UVA, Paseo Bel\'en, 7, 47011
Valladolid, Spain}

\begin{abstract}
Some dynamical spacetime solutions that describe ultra-compact objects are analysed. 
We show that by turning a static spacetime that describes a one-way wormhole into a dynamical time-dependent one, a singularity might arise in general. 
Several well-known wormhole solutions are studied, which are shown to become singular when promoted to a time-dependent wormhole. 
However, we show that one can construct a dynamical wormhole free of singularities. 
The corresponding Lagrangian that holds such dynamical spacetimes is also reconstructed, based on a two-scalar-field model free of ghosts. 
Energy conditions are also analysed, which might remain satisfied for a dynamical -singular- wormhole.
Finally, the evolution of the optical appearance of such dynamical objects is simulated and compared to static solutions. 
Some interesting features emerge that point to a way to measure the time evolution of dynamical objects by looking at the photon ring structure around the object. 
\end{abstract}

\maketitle

\section{Introduction}\label{Intr}

Although spherically symmetric and static spacetime, on the one hand, and spatially isometric and homogeneous cosmology, on the other, have been well studied for a long time, the time evolution of spacetime geometry with spatially non-trivial structures remains poorly understood, even when the spacetime is spherically symmetric. 
The McVittie metric~\cite{McVittie:1933zz} could be a first example where the black hole is embedded in the expanding universe. 
It has been shown in \cite{Nojiri:2020blr} that an arbitrary, dynamical (time-dependent), spherically symmetric spacetime can be realised by using Einstein's gravity coupled with two scalar fields. 
The model in \cite{Nojiri:2020blr} includes ghosts in general, but the ghosts can be excluded by using some constraints~\cite{Nojiri:2023dvf, Nojiri:2023zlp, Elizalde:2023rds, Nojiri:2023ztz, Nojiri:2024dde, Alencar:2025nik}. 
Furthermore, in \cite{Nashed:2024jqw, Katsuragawa:2024bwm}, it is pointed out that an arbitrary geometry in four dimensions can be realised using a non-linear $\sigma$ model whose target-space metric is identified with the Ricci curvature. 

Nevertheless, recently it became clear that there is a naked curvature singularity at the horizon~\cite{Modesto:2025cre} in the McVittie solution, 
In this paper, we show that this is a rather universal phenomenon, that is, the horizon embedding in a dynamical background becomes a naked singularity for a large class of models. 
We investigate how the singularity can be avoided and give general prescriptions to do so. 
We find that the geometry is strongly constrained. In particular, the radius of the horizon must be constant even if the black hole is embedded in a time-dependent spacetime. 

Then, by using such prescriptions, we construct some dynamical spacetimes where the wormhole is hidden inside the horizon(s), including the dynamical extension of the Simpson-Visser black bounce~\cite{Simpson:2018tsi}. 
We should note that in the static case, there are some numerical studies of a wormhole inside a stellar object~\cite{Dzhunushaliev:2011xx, Dzhunushaliev:2012ke, Dzhunushaliev:2013lna, Dzhunushaliev:2014mza, Aringazin:2014rva, Dzhunushaliev:2015sla, Dzhunushaliev:2016ylj, Dzhunushaliev:2022elv} and also of a wormhole inside a black hole has been well-investigated in Refs.~\cite{Simpson:2018tsi, Nojiri:2024dde}. 
The dynamical geometries in this paper can be constructed using two scalar fields without ghosts, following an extension of the formulation proposed in Refs.~\cite{Nojiri:2020blr, Nojiri:2023dvf, Nojiri:2023zlp, Elizalde:2023rds, Nojiri:2023ztz, Nojiri:2024dde, Alencar:2025nik}. 

In the dynamical spacetime, the separation of the energy-momentum tensor into the energy density and the pressures is not so trivial even if the spacetime is spherically symmetric, because there appear the four-velocity vector of the fluid, a unit space-like vector in the radial direction, and sometimes the heat flux vector. 
In this paper, we give the formulae for the separation explicitly. 
Then, we obtain the correct expressions of the energy density and the pressures in the radial and angular directions. 
By using those expressions, we discuss the energy conditions, which can be modified in the dynamical spacetime. 
By using the separation, we investigate the energy conditions and show that some conditions that are not satisfied in the static spacetime are satisfied in some dynamical cases. 

We also investigate the thermodynamics of the Simpson-Visser black bounce. In the static case, the thermodynamics is not changed from that of a Schwarzschild black hole. 
In the dynamical extension, we use the adiabatic approximation, where the time development of the system is very slow. 
We use the surface gravity to find the Hawking temperature $T_\mathrm{H}$, the ADM mass for the thermodynamical energy $E$, and the Bekenstein-Hawking entropy $\mathcal{S}_\mathrm{BH}$~\cite{Bekenstein:1972tm, Bekenstein:1973ur, Hawking:1975vcx}. 
We find that the obtained thermodynamical quantities are completely identical to those of the static case, that is, those in the Schwarzschild black hole. \\

Hence, in this paper we intend to show the following: 
\begin{enumerate}
\item When a static spacetime with a horizon is embedded in a dynamical (time-dependent) background, the horizon becomes a naked singularity, in general. 
\item Strong constraints on the geometry are obtained in order to avoid the above problem. 
Especially, the radius of the horizon must be a constant in time. 
\item Even in spherically symmetric spacetime, the relations between the energy-momentum tensor and the energy density and pressures are non-trivial and complicated. 
We explicitly find the expressions and discuss the energy conditions.
\item By using the technique of ray-tracing, we simulate the optical appearance of such objects when surrounded by an accretion disk and how the photon ring structure evolves with time. 
\end{enumerate}
For the above purposes, in the next section, we consider the following form of the metric in the dynamical (time-dependent) spherically symmetric spacetime 
\begin{align}
\label{ssWHmtrc}
ds^2 = - \e^{2\nu (t,r)} dt^2 + \e^{2\lambda (t,r)} dr^2 + \e^{2\xi (t,r)} \left( d\vartheta^2 + \sin^2\vartheta \, d\varphi^2 \right)\, , 
\end{align}
We define the metric $\bar{g}_{ij}$ of the unit 2-sphere by $\sum_{i,j=1,2} \bar{g}_{ij} dx^i dx^j = d\vartheta^2 + \sin^2\vartheta d\varphi^2$. 
Without loss of generality, we may choose $\e^{2\xi (t,r)}=r^2$ as shown in \cite{Nojiri:2020blr}, but when we consider dynamical or time-dependent wormholes, it is convenient to keep $\e^{2\xi (t,r)}$. 
The dynamical and spherically symmetric spacetime \eqref{ssWHmtrc} can be realised by Einstein's gravity coupled to two scalar fields, as in \cite{Nojiri:2020blr}. In Section~\ref{TSM}, we explicitly present the construction and show that ghosts can be excluded. In Section~\ref{Snglrprblm}, we show that the horizon embedded in the time-dependent spacetime becomes a naked singularity in rather general cases, and we show the conditions under which this problem could be avoided. 
In the dynamical and spherically symmetric spacetime, the general energy-momentum tensor has the following form, 
\begin{align}
\label{EMtensorGeneral}
T_{\mu\nu} =&\, \left(\rho + p_t \right) u_\mu u_\nu + p_t g_{\mu\nu} + \left( p_r - p_t\right) s_\mu s_\nu + q_\mu u_\nu + q_\nu u_\mu \, .
\end{align}
Here $\rho$ is the energy density, $p_r$ is the pressure in the radial direction, $p_t$ is the pressure in the angular or transverse direction, $u_\mu$ if the four-velocity vector of the fluid, 
The vector $s_\mu$ is a unit space-like vector in the radial direction and perpendicular to $u_\mu$, and $q_\mu$ is the heat flux vector, which is proportional to $s_\mu$. \\

The paper is organised as follows: in Section~\ref{EMtnsr}, we give the explicit form for $\rho$, $p_r$, $p_t$, $u_\mu$ and $s_\mu$ in terms of the energy-momentum tensor, which can be rewritten by using the curvature through the Einstein equation in the case of Einstein's gravity. 
No instabilities arise in the spacetime constructed within the two-scalar model in Section~\ref{TSM}, even in the case that there are violations of the energy conditions. 
In Section~\ref{EC}, the energy conditions are investigated and how they might be satisfied. 
We give an explicit example as in \cite{Katsuragawa:2025zcy}. 
In Section~\ref{SVbb}, as an example of the dynamical spacetime with a horizon, we extend the Simpson-Visser black bounce~\cite{Simpson:2018tsi} so that the spacetime evolves in time without a singularity at the horizon, and we discuss the energy conditions. 
In Section~\ref{TDdbbs}, we investigate the thermodynamics of the dynamical extension of the Simpson-Visser black bounce, and we show that the thermodynamics is not changed from that of the static Schwarzschild black hole. 
In Section~\ref{shadows-non-stationary-wormholes}, we analyse possible signatures reflecting the evolution of such dynamical spacetimes by simulating the optical appearance of these objects when surrounded by an accretion disk. 
Such images show a photon ring structure with a central brightness depression, called the ``shadow'' and defined, in general, as the central region formed by photons captured by the unstable circular orbit, the so-called photon sphere~\cite{Bronzwaer:2021lzo, Gralla:2019xty, Perlick:2021aok}. 
This type of analysis has become very popular as the images of the central objects located at M87* and Sgr A* might reveal fundamental properties of the spacetime structure around them~\cite{EventHorizonTelescope:2019dse, EventHorizonTelescope:2022wkp, EventHorizonTelescope:2022xqj, EventHorizonTelescope:2021dqv}. 
Most of the previous analyses are focused on static spacetimes, motivated mainly to find clear fingerprints in the images that show deviations from the Kerr paradigm~\cite{Staelens:2023jgr, Guerrero:2021ues, Kocherlakota:2023qgo, Olmo:2023lil, Olmo:2025ctf}. 
Then, extending shadow calculations to non-stationary solutions is a fundamental and natural step to better understand the real reconstruction of the images, where real objects show a great variability and to extract those features that depend on variations of the spacetime metric itself in order to distinguish them from those time-dependent magnitudes that belong to the variability of the accretion disk. 
Hence, we explore how varying the metric may induce changes in the observed luminosity. 
Finally, Section~\ref{conclusions} gathers the summary and conclusions of the paper.

\section{Two scalar fields model}\label{TSM}

In this section, we show that the spacetime given in \eqref{ssWHmtrc} can be realised by Einstein's gravity coupled to two scalar fields, as shown in \cite{Nojiri:2020blr}. 
Let us first consider Einstein's gravity with two scalar fields $\phi$ and $\chi$, whose action is described by 
\begin{align}
\label{I8}
S_{( \mathrm{GR} \phi\chi)} = \int d^4 x \sqrt{-g} & \left[ \frac{R}{2\kappa^2} 
 - \frac{1}{2} \, A (\phi,\chi) \partial_\mu \phi \partial^\mu \phi - B (\phi,\chi) \, \partial_\mu \phi \partial^\mu \chi 
 - \frac{1}{2} \, C (\phi,\chi) \partial_\mu \chi \partial^\mu \chi - V (\phi,\chi)\right] \, .
\end{align}
Here $g$ is the determinant of the metric $g_{\mu\nu}$, $V(\phi, \chi)$ is the potential of the scalar fields $\phi$ and $\chi$ and the coefficient functions 
$A$, $B$ and $C$ depend on the scalars $\phi$ and $\chi$. 
The energy-momentum tensor is given by 
\begin{align}
\label{I9}
T^{(\phi\chi)}_{\mu\nu} =& g_{\mu\nu} \left[ 
 - \frac{1}{2}\, A (\phi,\chi) \partial_\rho \phi \partial^\rho \phi 
 - B (\phi,\chi) \partial_\rho \phi \partial^\rho \chi 
 - \frac{1}{2} \, C (\phi,\chi) \partial_\rho \chi \partial^\rho \chi - V 
(\phi,\chi)\right] \nonumber \\
& + A (\phi,\chi) \partial_\mu \phi \partial_\nu \phi 
+ B (\phi,\chi) \left( \partial_\mu \phi \partial_\nu \chi 
+ \partial_\nu \phi \partial_\mu \chi \right) 
+ C (\phi,\chi) \partial_\mu \chi \partial_\nu \chi \, , 
\end{align}
and the contracted Bianchi identities have the following forms, 
\begin{align}
\label{Bianchiphi}
0 =& \, \frac{A_\phi}{2} \partial_\mu \phi \partial^\mu \phi 
+ A \nabla^\mu \partial_\mu \phi + A_\chi \partial_\mu \phi \partial^\mu 
\chi + \left( B_\chi - \frac{1}{2} \, C_\phi \right)\partial_\mu \chi 
\partial^\mu \chi + B \nabla^\mu \partial_\mu \chi - V_\phi \, ,\\
%&\nonumber \\ 
\label{Bianchichi}
0 =& \left( - \frac{1}{2} \, A_\chi + B_\phi \right) \partial_\mu \phi 
\partial^\mu \phi 
+ B \nabla^\mu \partial_\mu \phi 
+ \frac{1}{2} \, C_\chi \partial_\mu \chi \partial^\mu \chi + C 
\nabla^\mu \partial_\mu \chi 
+ C_\phi \partial_\mu \phi \partial^\mu \chi - V_\chi\, .
\end{align}
Here $A_\phi \equiv \partial A(\phi,\chi)/\partial \phi$, {\em etc}. 
We may identify $\phi$ and $\chi$ as temporal and radial coordinates, respectively, 
\begin{align}
\label{TSBH1}
\phi=t\, , \quad \chi=r\, .
\end{align}
We should note that this assumption \eqref{TSBH1} does not lead to any loss of generality. 
This is because for the spherically symmetric solutions~\eqref{ssWHmtrc} of the theory~\eqref{I8}, in general $\phi$ and $\chi$ depend on both coordinates $t$ and $r$. 
If any solution is given, we can determine the $t$- and $r$-dependence of $\phi$ and $\chi$ so that $\phi$ and $\chi$ are given by specific functions $\phi(t,r)$, $\chi(t,r)$. 
Then we can redefine the scalar fields to replace $t$ and $r$ with new scalar fields, say, $\bar{\phi}$ and $\bar{\chi}$ with $\phi(t,r)\to \phi(\bar{\phi}, \bar{\chi})$ and $\chi(t,r) \to \chi(\bar{\phi}, \bar{\chi})$. 
This allows us to identify the new fields with $t$ and $r$ as in~\eqref{TSBH1}), $\bar{\phi}\to \phi=t$ and $\bar{\chi} \to \chi= r$. 
The change of variables $\left( \phi , \chi\right) \rightarrow \left( \bar{\phi} , \bar{\chi} \right)$ can be always absorbed into redefinitions of $A$, $B$, $C$ and $V$ in the action~\eqref{I8}. 
Hence, the assumption~\eqref{TSBH1} does not lead to any loss of generality. 

By using \eqref{curvaturesWH}, we find that the non-vanishing components of the Einstein tensor $G_{\mu\nu}=R_{\mu\nu} - \frac{1}{2} g_{\mu\nu}$ are the $\left( t,t \right)$, 
$\left(r,r\right)$, $\left(i,j\right)$ and $\left(t,r \right)$ components. 
Even for the energy-momentum tensor in \eqref{I9}, the non-vanishing components are the $\left( t,t \right)$, $\left(r,r\right)$, $\left(i,j\right)$, and $\left(t,r \right)$ components. 
Then the Einstein equation gives, 
\begin{align}
\label{TSBH2}
\frac{1}{\kappa^2} \left( R_{tt} + \frac{1}{2}\e^{2\nu} R \right)
=&\, - \e^{2\nu} \left[ \frac{1}{2} A \e^{-2\nu} - \frac{1}{2} C \e^{-2\lambda} - V \right] + A 
= - \e^{2\nu} \left( - \frac{A}{2} \e^{-2\nu} - \frac{C}{2} \e^{-2\lambda} - V \right) \, ,\\
\label{TSBH3}
\frac{1}{\kappa^2} \left( R_{rr} - \frac{1}{2}\e^{2\lambda} R \right)
=&\, \e^{2\lambda} \left[ \frac{1}{2} A \e^{-2\nu} - \frac{1}{2} C \e^{-2\lambda} - V \right] + C 
= \e^{2\lambda} \left( \frac{A}{2} \, \e^{-2\nu} + \frac{C}{2} \e^{-2\lambda} - V \right) \, ,\\
\label{TSBH4}
\frac{1}{\kappa^2} R_{tr} = \frac{1}{\kappa^2} R_{rt} =&\, B \, . \\
\label{TSBH5}
\frac{1}{\kappa^2} \left( \frac{1}{2} {\bar g}^{ij} R_{ij} - \frac{1}{2} \e^{2\xi} R \right) =&\, 
\e^{2\xi} \left( \frac{A}{2} \e^{-2\nu} - \frac{C}{2} \e^{-2\lambda} - V \right) \, . 
\end{align}
Due to spherical symmetry and staticity, the other components of the Einstein equation are trivially satisfied. 
Equations~\eqref{TSBH2}-\eqref{TSBH5} can be solved with respect to $A$, $B$, $C$ and $V$, obtaining the inverse relations 
\begin{align}
\label{WHA}
A =&\, \frac{1}{\kappa^2} \left( R_{tt} + \frac{1}{2} \e^{2\nu-2\xi} {\bar g}^{ij} R_{ij} \right) \, , \\
\label{WHB}
B =&\, \frac{1}{\kappa^2} R_{tr} = \frac{1}{\kappa^2} R_{rt} \, , \\
\label{WHC}
C =&\, \frac{1}{\kappa^2} \left( R_{rr} - \frac{1}{2} \e^{2\lambda-2\xi} {\bar g}^{ij} R_{ij} \right) \, , \\
\label{WHV}
V =&\, \frac{1}{2\kappa^2} \left( \e^{-2\nu}R_{tt} - \e^{-2\lambda} R_{rr} + R\right) \, .
\end{align}
Because the r.h.s. of Eqs.~\eqref{WHA}-\eqref{WHV} are functions of $t$ and $r$, by replacing $\left( t, r\right) $ with $\left( \phi , \chi \right) $ in these r.h.s., $A$, $B$, $C$ and $V$ are given by the functions of $\left( \phi , \chi \right)$. 
Conversely, if $A$, $B$, $C$ and $V$ are prescribed, the spherically symmetric configurations~\eqref{ssWHmtrc} corresponding to arbitrary functions $\nu$, $\lambda$ and $\xi$ are solutions of the model. 

It should be noted that the functions $A$ and/or $C$ often become negative, and therefore $\phi$ and/or $\chi$ are ghosts. 
The ghosts can be eliminated by imposing constraints by introducing the Lagrange multiplier fields $\lambda_\phi$ and $\lambda_\chi$ and modifying the action \eqref{I8} $S_{\mathrm{GR} \phi\chi} \to S_{\mathrm{GR} \phi\chi} + S_\lambda$. 
Here the additional term $S_\lambda$ is given by~\cite{Nojiri:2023dvf, Nojiri:2023zlp, Elizalde:2023rds, Nojiri:2023ztz, Nojiri:2024dde}
\begin{align}
\label{lambda1}
S_\lambda = \int d^4 x \sqrt{-g} \left[ \lambda_\phi \left( \e^{-2\nu(t=\phi, r=\chi)} \partial_\mu \phi \partial^\mu \phi + 1 \right)
+ \lambda_\chi \left( \e^{-2\lambda(t=\phi, r=\chi)} \partial_\mu \chi \partial^\mu \chi - 1 \right) \right] \, .
\end{align}
By the variation of the action $S_\lambda$ with respect to $\lambda_\phi$ and $\lambda_\chi$, the following constraints are obtained, 
\begin{align}
\label{lambda2}
0 = \e^{-2\nu(t=\phi, r=\chi)} \partial_\mu \phi \partial^\mu \phi + 1 \, , \quad
0 = \e^{-2\lambda(t=\phi, r=\chi)} \partial_\mu \chi \partial^\mu \chi - 1 \, ,
\end{align}
which is consistent with the assumption \eqref{TSBH1}.

Due to the constraints from Eq.~\eqref{lambda2}, the scalar fields $\phi$ and $\chi$ become non-dynamical, and therefore the fluctuations of $\phi$ and $\chi$ from the background \eqref{TSBH1} do not propagate. 
The fluctuations are now written as follows, 
\begin{align}
\label{pert1}
\phi=t + \delta \phi \, , \quad \chi=r + \delta \chi\, .
\end{align}
By using Eq.~\eqref{lambda2}, we obtain 
\begin{align}
\label{pert2}
\partial_t \left( \e^{-2\nu(t,r)} \delta \phi \right) = \partial_r \left( \e^{-2\lambda(t,r)} \delta \chi \right) = 0\, .
\end{align}
The equation~\eqref{pert2} tells us that, by imposing the initial condition $\delta\phi=0$ and by imposing the boundary condition $\delta\chi\to 0$ when $r\to \infty$, we can find that both of $\delta \phi$ and $\delta \chi$ vanish in the whole spacetime, $\delta\phi=0$ and $\delta\chi=0$. 
This tells us that both $\phi$ and $\chi$ are non-dynamical or frozen degrees of freedom. 

As partially or completely shown in Refs.~\cite{Nojiri:2023dvf, Nojiri:2023zlp, Elizalde:2023rds, Nojiri:2023ztz}, even in the model given by the modified action $S_{\mathrm{GR} \phi\chi} + S_\lambda$, $\lambda_\phi=\lambda_\chi=0$ consistently appear as a solution and therefore any solution in the original action~\eqref{I8} is a solution even for the modified model with the action $S_{\mathrm{GR} \phi\chi} + S_\lambda$. 

\section{Singularity problems}\label{Snglrprblm}

We consider exotic objects, such as a wormhole whose throat is hidden by a black hole. 
The wormhole inside a stellar object has been studied in Refs.~\cite{Dzhunushaliev:2011xx, Dzhunushaliev:2012ke, Dzhunushaliev:2013lna, Dzhunushaliev:2014mza, Aringazin:2014rva, Dzhunushaliev:2015sla, Dzhunushaliev:2016ylj, Dzhunushaliev:2022elv} numerically within Einstein's gravity, and the geometry of the wormhole inside a black hole has been well-investigated in Refs.~\cite{Simpson:2018tsi, Nojiri:2024dde}. 

As a solution describing a dynamical or time-dependent black hole, a well-known exact solution of the Einstein equation is the McVittie metric~\cite{McVittie:1933zz}, 
Recently, however, it has been found that even for the McVittie solution, there is a naked curvature singularity at the horizon~\cite{Modesto:2025cre}. 
In this section, we show that this phenomena that the horizon becomes a naked singularity seem to be rather general, and we find how this problem could be avoided. 

At the horizon, where $\e^{2\nu}$ vanishes, the curvatures in \eqref{curvaturesWH} are singular in general. 
In order to avoid the singularity, we need to require that $ \ddot\lambda + \left( \dot\lambda - \dot\nu \right) \dot\lambda$ and $\ddot\xi + \left(\dot\xi - \dot\nu \right) \dot \xi$ vanish at the horizon, 
Then we require, 
\begin{align}
\label{r1}
\ddot\lambda + \left( \dot\lambda - \dot\nu \right) \dot\lambda = \Lambda_0 \left(t, r\right) \e^{2\nu} \, , \quad 
\ddot\xi + \left(\dot\xi - \dot\nu \right) \dot \xi = \Xi_0 \left(t, r\right) \e^{2\nu} \, .
\end{align}. 
Here $\Lambda_0 \left(t, r\right)$ and $\Xi_0 \left(t, r\right)$ are smooth functions of $t$ and $r$. 
The equations in \eqref{r1} can be rewritten as 
\begin{align}
\label{r2}
\frac{\partial}{\partial t} \left( \e^{-\nu} \frac{\partial \e^\lambda}{\partial t} \right) = \e^{\lambda - \nu} \Lambda_0 \left(t, r\right) \e^{2\nu} \, , \quad 
\frac{\partial}{\partial t} \left( \e^{-\nu} \frac{\partial \e^\xi}{\partial t} \right) = \e^{\xi - \nu} \Xi_0 \left(t, r\right) \e^{2\nu} \, .
\end{align}
In order that the curvature is not singular at the horizon, we require that $\e^{2\lambda + 2\nu}$ is not singular and does not vanish at the horizon. 
Therefore we may absorb $\e^{\lambda + \nu}$ into the redefinition of $\Lambda_0$. 
On the other hand, $\e^{2\xi}$ should not be singular and does not vanish at the horizon, and therefore, we may absorb $\e^\xi$ into the redefinition of $\Xi_0$. 
Then we rewrite \eqref{r2} as follows, 
\begin{align}
\label{r3}
\frac{\partial}{\partial t} \left( \e^{-\nu} \frac{\partial \e^\lambda}{\partial t} \right) = \Lambda \left(t, r\right) \, , \quad 
\frac{\partial}{\partial t} \left( \e^{-\nu} \frac{\partial \e^\xi}{\partial t} \right) = \Xi \left(t, r\right) \e^\nu \, , \quad 
\Lambda \left(t, r\right) \equiv \e^{\lambda - \nu} \Lambda_0 \left(t, r\right) \, , \quad 
\Xi \left(t, r\right) \equiv \e^\xi \Xi \left(t, r\right) \, ,
\end{align}
which can be solved as, 
\begin{align}
\label{r4}
\e^{\lambda (t,r)} = \int^t dt_1 \e^{\nu\left(t_1, r\right)} \int^{t_1} dt_2 \Lambda \left(t_2, r\right) \, , \quad 
\e^{\xi (t,r)} = \int^t dt_1 \e^{\nu\left(t_1, r\right)} \int^{t_1} dt_2 \Xi \left(t_2, r\right) \e^{\nu\left(t_2, r\right)} \, .
\end{align}
By integrating with respect to $t_1$ and $t_2$, there appear four functions that depend only on $r$, which correspond to the integration constants. 
Eq.~\eqref{r4} provides a general class of spherically symmetric, time-dependent black hole or wormhole geometries. 
An additional strong requirement is that $\e^{2\lambda}$ becomes singular when $\e^{2\nu}$ vanishes, so that $\e^{2\lambda + 2\nu}$ should be finite and does not vanish, nor curvature singularity appears at the horizon. 
Because the integrand in the r.h.s. of the first equation in \eqref{r4} is regular, the singularity should appear as constants of the integrations for $t_1$ and $t_2$. 
This requires that the horizon radius given by $r$ should not depend on time. 
This does not mean, however, that the physical radius of the horizon is necessarily constant because the physical angular radius is given by $\e^\xi$. 
If we regard $r$ as a function of $t$ and $\xi$, the horizon radius may be time-dependent. 

In summary, in order that curvature singularities do not appear at the horizon(s), we require, 
\begin{itemize}
\item At the horizon, $ \ddot\lambda + \left( \dot\lambda - \dot\nu \right) \dot\lambda$, $\ddot\xi + \left(\dot\xi - \dot\nu \right) \dot \xi$ vanish. ,
\item At the horizon, $\e^{2\lambda + 2\nu}$ is not singular nor does not vanish.
\end{itemize}
Then the conditions for the geometry in the time-dependent spacetime are 
\begin{enumerate}
\item There is a radial coordinate where the horizon radius in terms of the coordinate does not depend on time. 
\item When we denote the above radial coordinate by $r$ and any time coordinate by $t$, we can choose the metric form as in \eqref{ssWHmtrc}. 
\item In the metric choice \eqref{ssWHmtrc}, $\e^{\lambda (t,r)}$ and $\e^{\xi (t,r)}$ can be given by \eqref{r4}. 
\item In \eqref{r4}, constant(s) of the integration for $\e^{\lambda (t,r)}$ must be singular so that $\e^{2\lambda + 2\nu}$ should be finite and does not vanish. 
\end{enumerate}
The above conditions, of course, do not exclude possible singularities in the past or future universe and the black hole singularity at the origin. 

When the above conditions are satisfied, although the real radius of the horizon is given by the areal radius $\e^\xi$, we find that the horizon radius is time-independent as we now show: 
The conditions require that near the horizon, $\e^{\nu}$ should be expanded as $\e^\nu = \nu_0 (t) \sqrt{r - r_\mathrm{h}} + \mathcal{O} \left( r - r_\mathrm{h} \right)$ with a function $\nu_0 (t)$, which only depends on $t$. 
Here $r_\mathrm{h}$ is the radius of the horizon in terms of $r$, and it does not depend on time under the conditions. 
Then the second equation in \eqref{r4} tells
\begin{align}
\label{r4B}
\e^{\xi (t,r)} = \e^{\xi_0(r)} + \left( r - r_\mathrm{h} \right) \int^t dt_1 \nu_0 \left(t_1\right) \int^{t_1} dt_2 \nu_0 \left(t_2\right) \Xi \left(t_2, r\right) 
+ \mathcal{O} \left( \left(r - r_\mathrm{h}\right)^\frac{3}{2} \right) \, .
\end{align}
Here $\e^{\xi_0(r)}$ appears as a constant of the integration with respect to $t_1$ and $t_2$. 
Then, in terms of the areal radius $\e^\xi$, the radius of the horizon is given by $\e^\xi=\e^{\xi_0\left( r= r_\mathrm{h} \right)}$, which does not depend on the time coordinate $t$. 

\subsection{Embedding of the black hole in the FLRW universe}

We now consider the embedding of the Schwarzschild-type black hole in the Friedmann-Lema\^{i}tre-Robertson-Walker (FLRW) universe. 
An example, which satisfies the conditions, is given by 
\begin{align}
\label{Schwrztyp}
ds^2 = - \left( 1 - \frac{2M}{r} \right) dt^2 + \frac{1}{1-\frac{2M}{r}}\left\{ \frac{2M}{r} + a(t) \left( 1 - \frac{2M}{r} \right) \right\}^2 dr^2 
+ \left\{ a(t)\left( r - 2M \right) + 2M\right\}^2 \sum_{i,j=1,2} \bar{g}_{ij} dx^i dx^j \, ,
\end{align}
that is, 
\begin{align}
\label{Schwrztyp2}
\e^\lambda = \sqrt{\frac{\frac{2M}{r}}{1 - \frac{2M}{r}}} + a(t) \left( 1 - \frac{2M}{r} \right)^\frac{1}{2} \, , \quad 
\e^\xi = a(t)\left( r - 2M \right) + 2M \, ,
\end{align}
which gives 
\begin{align}
\label{Schwrztyp3}
\Lambda \left(t, r\right) = \ddot a (t) \, , \quad \Xi \left(t, r\right) = \ddot a (t) r \, .
\end{align}
When $r\to \infty$, the metric in \eqref{Schwrztyp} behaves as a standard FLRW universe. 
\begin{align}
\label{Schwrztyp4}
ds^2 \sim - dt^2 + a(t)^2 dr^2 + a(t)^2 r^2 \sum_{i,j=1,2} \bar{g}_{ij} dx^i dx^j \, .
\end{align}
On the other hand, near the horizon $r\sim 2M$, the metric in \eqref{Schwrztyp} reduces to the Schwarzschild metric, 
\begin{align}
\label{Schwrztyp5}
ds^2 = - \left( 1 - \frac{2M}{r} \right) dt^2 + \frac{1}{1 - \frac{2M}{r}} dr^2 + \left( 2M \right)^2 \sum_{i,j=1,2} \bar{g}_{ij} dx^i dx^j \, .
\end{align}
Although the real radius of the horizon is given by the areal radius $\e^\xi$, we find $r=\e^\xi=2M$ at the horizon, and therefore, the horizon radius is time-independent. 

\subsection{Models of wormhole}

As a first model, we consider the Hayward black hole~\cite{Hayward:2005gi}-type metric, 
\begin{align}
\label{Hayward}
\e^{2\nu}=\e^{-2\lambda}=1 - \frac{2Mr^2}{\left|r\right|^3 + 2M\gamma^2} \, .
\end{align}
In the original metric, $\left|r\right|^3$ was $r^3$, but we like to assume that $r$ runs from $-\infty$ to $+\infty$ because we will consider the wormhole spacetime. 
If we choose $r^3$ instead of $\left|r\right|^3$ in \eqref{Hayward}, a singularity appears when $r^3 + 2M\gamma^2=0$. 
We should note that $\e^{2\nu}=\e^{-2\lambda}$ and its first and second derivatives are continuous at $r=0$, and therefore the curvatures are also continuous. 
If we need to require that the higher derivatives of $\e^{2\nu}=\e^{-2\lambda}$ are also continuous, we may replace $\left|r\right|^3 + 2M\gamma^2$ in \eqref{Hayward} with $\sqrt{r^6 + 4M^2\gamma^4}$. 
The choice of $\e^{2\lambda}$ is given by choosing $\Xi \left(t, r\right)=0$ in \eqref{r4} and the `constant' of the integration with respect $t$-integration to be, $\e^\lambda=\left( 1 - \frac{2Mr^2}{r^3 + 2M\gamma^2} \right)^{-\frac{1}{2}}$. 
As we discussed, the Hayward-type choice in \eqref{Hayward} does not mean that the spacetime is time-independent because the angular radius is given by $\e^\xi$. 

When $r>0$, because 
\begin{align}
\label{min}
\left( \e^{2\nu} \right)' = \frac{2Mr \left( r^3 - 4M\gamma^2\right)}{\left( r^3 + 2M\gamma^2 \right)^2}\, ,
\end{align}
$\e^{2\nu}$ has a minimum when $r=\left( 4M\gamma^2 \right)^\frac{1}{3}$. 
The minimum value is given by 
\begin{align}
\label{numin}
\left. \e^{2\nu} \right|_{r=\left( 4M\gamma^2 \right)^\frac{1}{3}} 
= 1 - \frac{2^\frac{5}{3}M}{3\left( 2M\gamma^2 \right)^\frac{1}{3}}\, .
\end{align}
Then we find the following, 
\begin{enumerate}
\item In the case that the following inequality is satisfied,
\begin{align}
\label{condition1}
\frac{2^\frac{5}{3}M}{3\left( 2M\gamma^2 \right)^\frac{1}{3}}<1\, ,
\end{align}
$\e^{2\nu}$ is always positive. 
This metric corresponding to \eqref{Hayward} could describe a kind of gravastar, which was proposed in \cite{Mazur:2001fv} as an alternative to the black hole.
\item On the other hand, when
\begin{align}
\label{condition2}
\frac{2^\frac{5}{3}M}{3\left( 2M\gamma^2 \right)^\frac{1}{3}}>1\, ,
\end{align}
$\e^{2\nu}$ vanishes twice. 
These two zeros correspond to the outer and inner horizons, and therefore the metric corresponds to the regular black hole in \cite{Hayward:2005gi} in the case of $\e^\xi=r$. 
The horizon radii $r_\pm$ are given by 
\begin{align}
\label{rpm}
r_\pm =\frac{M^\frac{2}{3}\left[ \left\{8M^2-27\gamma^2 \pm 3\lambda\sqrt{81\gamma^2-48M^2} \right\}^\frac{1}{3}
+2M^\frac{2}{3} \right] \left\{8M^2-27\gamma^2 \pm 3\lambda\sqrt{81\gamma^2-48M^2} \right\}^\frac{1}{3} +4M^2}
{3M^\frac{1}{3} \left\{ 8M^2-27\gamma^2 \pm 3\lambda\sqrt{81\gamma^2-48M^2} \right\}^\frac{1}{3} }\,.
\end{align}
\item When
\begin{align}
\label{condition3}
\frac{2^\frac{5}{3}M}{3\left( 2M\gamma^2 \right)^\frac{1}{3}}=1\, ,
\end{align}
the radii of the two horizons become identical with each other and constitute the degenerate horizon, and therefore, it corresponds to the extremal limit of the black hole. 
Such an extremal limit may generate the Nariai-type geometry~\cite{Nariai:1999iok}. 
\end{enumerate}

The second equation in \eqref{r4} gives, 
\begin{align}
\label{xi1}
\e^{\xi\left(t,r\right)} = \Xi_1 (r) + \e^{2\nu(r)} \Xi_2(r,t)= \Xi_1 (r) + \left( 1 - \frac{2Mr^2}{r^3 + 2M\gamma^2} \right) \Xi_2(r,t)
\, , \quad \Xi_2(r,t) \equiv \int^t dt_1 \int^{t_1} dt_2 \Xi \left(t_2, r\right) \, .
\end{align}
$\Xi_1$ appears as a `constant' of the integration with respect to the $t$-integration. 
We should note that $\Xi_1 (r)$ can be an arbitrary function of $r$ and $\Xi_2(r,t)$ can also be an arbitrary function of $t$ and $r$ as long as a smooth function. 
\begin{itemize}
\item {\bf static case}
First of all, we may consider a static model.
\begin{align}
\label{stmdl}
\Xi_1 (r) = \sqrt{r^2 +{r_\mathrm{throat}}^2}\, , \quad \Xi_2(r,t)= 0\quad \mbox{that is} \quad \e^{2\xi} = r^2 +{r_\mathrm{throat}}^2 \, .
\end{align}
Here we assume $r$ runs from $-\infty$ to $+\infty$. 
If the condition~\eqref{condition1} is satisfied, the spacetime is a wormhole whose throat radius is given by $r_\mathrm{throat}$. 
On the other hand, if the condition~\eqref{condition2} is satisfied, the geometry of the wormhole inside a black hole is realised as in Refs.~\cite{Simpson:2018tsi, Nojiri:2024dde}. 
The wormhole throat connects two universes, one corresponding to $r>0$ and the other to $r<0$. 
\item {\bf dynamical case 1} As an example, we consider the following,
\begin{align}
\label{dynmclmdl}
\Xi_1 (r) = \sqrt{r^2 +{r_\mathrm{throat}}^2}\, , \quad \Xi_2=\left( 1 - \frac{2Mr^2}{r^3 + 2M\gamma^2} \right)^{-1} \Xi_{02}(t) \quad \mbox{that is}\quad 
\e^{\xi} = \sqrt{r^2 +{r_\mathrm{throat}}^2} + \Xi_{02}(t) \, .
\end{align}
Here we assume the condition~\eqref{condition1} and therefore the function $\left( 1 - \frac{2Mr^2}{r^3 + 2M\gamma^2} \right)^{-1}$ does not have any pole and regular. 
In the model~\eqref{dynmclmdl}, the throat radius ${\bar r}_\mathrm{throat}$ is time-dependent 
\begin{align}
\label{thdynml}
{\bar r}_\mathrm{throat} = r_\mathrm{throat} + \Xi_{02}(t) \, .
\end{align}
If the throat radius ${\bar r}_\mathrm{throat}$ becomes very large, the universe might be swallowed by the wormhole. 
\item {\bf dynamical case 2} Another example is given by
\begin{align}
\label{dynmclmdl2}
\Xi_1 (r) = \sqrt{r^2 +{r_\mathrm{throat}}^2}\, , \quad \Xi_2= \Xi_{02}(t) \quad \mbox{that is }\quad 
\e^{\xi} = \sqrt{r^2 +{r_\mathrm{throat}}^2} + \left( 1 - \frac{2Mr^2}{\left|r\right|^3 + 2M\gamma^2} \right)\Xi_{02}(t) \, .
\end{align}
Eq.~\eqref{numin} tells 
\begin{align}
1 \geq \e^{2\nu} = 1 - \frac{2Mr^2}{\left|r\right|^3 + 2M\gamma^2} \geq 1 - \frac{2^\frac{5}{3}M}{3\left( 2M\gamma^2 \right)^\frac{1}{3}}\, .
\end{align}
If we choose $\Xi_{02}(t)$ so that 
\begin{align}
\label{ex2}
r_\mathrm{throat} + \left\{ 1 - \frac{2^\frac{5}{3}M}{3\left( 2M\gamma^2 \right)^\frac{1}{3}} \right\} \Xi_{02}(t) >0\, ,
\end{align}
we obtain the spacetime where there is a wormhole throat inside the black hole as in \cite{Simpson:2018tsi, Nojiri:2024dde}, but now we obtain a dynamical spacetime, where the horizon radii and the throat radius vary in time. 
\end{itemize}

We now show that when the two universes are separated or merged by the wormhole throat, the curvature singularity always appears. 
If the throat radius is negative $r_\mathrm{throat}<0$ in general, there is no throat and $\e^\xi$ can be negative when $r\to 0$. 
The expression of the scalar curvature $R$ in \eqref{curvaturesWH} indicates that there is a singularity when $\e^{2\xi}$ vanishes. 
If the condition~\eqref{condition1} is satisfied, the singularity is naked. 
In order to avoid the singularity, it is necessary that $1 + \e^{-2\nu+2\xi} {\dot\xi}^2 - \e^{-2\lambda+2\xi} {\xi'}^2$ should vanish when $\e^{2\xi}$ vanishes. 
In the case of \eqref{dynmclmdl}, we find 
\begin{align}
\label{v1}
1 &\, + \e^{-2\nu+2\xi} {\dot\xi}^2 - \e^{-2\lambda+2\xi} {\xi'}^2 \nonumber \\
=&\, 1 + \frac{\frac{r^2}{r^2 +{r_\mathrm{throat}}^2} - \left( 1 - \frac{2Mr^2}{r^3 + 2M\gamma^2} \right)^2 {{\dot \Xi}_{02} (t)}^2
}{\left( 1 - \frac{2Mr^2}{r^3 + 2M\gamma^2} \right)\left(\sqrt{r^2 +{r_\mathrm{throat}}^2} + \Xi_{02}(t)\right)^2} \, .
\end{align}
When $\e^\xi$ vanishes, $\sqrt{r^2 +{r_\mathrm{throat}}^2} + \Xi_{02}(t)$ vanishes or $r^2 = - {r_\mathrm{throat}}^2 + {\Xi_{02}(t)}^2$ and $\Xi_{02}(t)<0$. 
Therefore we require 
\begin{align}
\label{rq1}
0 =&\, \left. \frac{r^2}{r^2 +{r_\mathrm{throat}}^2} - \left( 1 - \frac{2Mr^2}{r^3 + 2M\gamma^2} \right)^2 {{\dot \Xi}_{02} (t)}^2 \right|_{r^2 = - {r_\mathrm{throat}}^2 + {\Xi_{02}(t)}^2} \nonumber \\
=&\, 1 - \frac{{r_\mathrm{throat}}^2}{{\Xi_{02}(t)}^2} - \left( 1 - \frac{2M\left( - {r_\mathrm{throat}}^2 + {\Xi_{02}(t)}^2\right)}{\left( - {r_\mathrm{throat}}^2 + {\Xi_{02}(t)}^2\right)^\frac{3}{2} + 2M\gamma^2} \right)^2 {{\dot \Xi}_{02} (t)}^2 \, , \\
\label{rq2}
0 =&\, \left. \frac{\partial}{\partial r} \left\{\frac{r^2}{r^2 +{r_\mathrm{throat}}^2} - \left( 1 - \frac{2Mr^2}{r^3 + 2M\gamma^2} \right)^2 {{\dot \Xi}_{02} (t)}^2 \right\} \right|_{r^2 = - {r_\mathrm{throat}}^2 + {\Xi_{02}(t)}^2} 
\nonumber \\
=&\, 2 \sqrt{- {r_\mathrm{throat}}^2 + {\Xi_{02}(t)}^2} \nonumber \\
&\, \times \left\{ \frac{{r_\mathrm{throat}}^2}{{\Xi_{02}(t)}^4} + \frac{-2M\left( - {r_\mathrm{throat}}^2 + {\Xi_{02}(t)}^2\right)^\frac{3}{2} + 8M^2\gamma^2}
{\left(\left( - {r_\mathrm{throat}}^2 + {\Xi_{02}(t)}^2\right)^\frac{3}{2} + 2M\gamma^2\right)^2} \left( 1 - \frac{2M\left( - {r_\mathrm{throat}}^2 + {\Xi_{02}(t)}^2 \right)}
{\left( - {r_\mathrm{throat}}^2 + {\Xi_{02}(t)}^2\right)^\frac{3}{2} + 2M\gamma^2} \right) 
 {{\dot \Xi}_{02} (t)}^2 \right\} \, .
\end{align}
In order that Eq.~\eqref{rq1} is compatible with \eqref{rq2}, we find 
\begin{align}
\label{rq3}
 - \frac{1 - \frac{{r_\mathrm{throat}}^2}{{\Xi_{02}(t)}^2}}{ \left( 1 - \frac{2M\left( - {r_\mathrm{throat}}^2 + {\Xi_{02}(t)}^2\right)}
{\left( - {r_\mathrm{throat}}^2 + {\Xi_{02}(t)}^2\right)^\frac{3}{2} + 2M\gamma^2} \right)^2}
= \frac{\frac{{r_\mathrm{throat}}^2}{{\Xi_{02}(t)}^4}}{\frac{-2M\left( - {r_\mathrm{throat}}^2 + {\Xi_{02}(t)}^2\right)^\frac{3}{2} + 8M^2\gamma^2}
{\left(\left( - {r_\mathrm{throat}}^2 + {\Xi_{02}(t)}^2\right)^\frac{3}{2} + 2M\gamma^2\right)^2} \left( 1 - \frac{2M\left( - {r_\mathrm{throat}}^2 + {\Xi_{02}(t)}^2 \right)}
{\left( - {r_\mathrm{throat}}^2 + {\Xi_{02}(t)}^2\right)^\frac{3}{2} + 2M\gamma^2} \right)} \, .
\end{align}
Because Eq.~\eqref{rq3} is an algebraic equation, $\Xi_{02}(t)$ must be a time-independent constant. 
This tells us that when the two universes are separated or merged by the wormhole throat, the curvature singularity always appears, which might be consistent because the throat becomes a conical singularity when the throat is pinched off. 

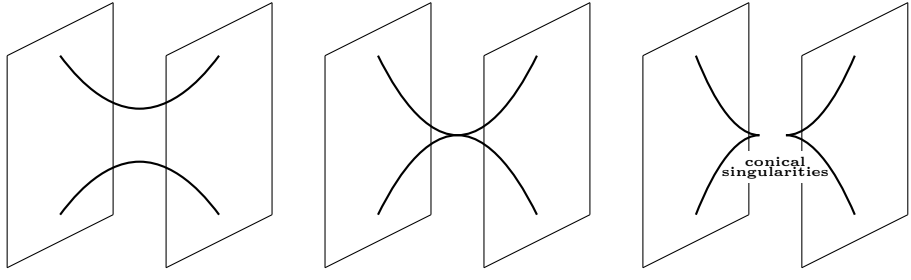
\begin{figure}[h]
\begin{center}

\begin{tikzpicture}[x=0.7cm, y=0.7cm]

\draw[thin] (0,0) -- (0,4); 
\draw[thin] (2,1) -- (2,5); 
\draw[thin] (0,0) -- (2,1); 
\draw[thin] (0,4) -- (2,5);

\draw[thin] (3,0) -- (3,4); 
\draw[thin] (5,1) -- (5,5); 
\draw[thin] (3,0) -- (5,1); 
\draw[thin] (3,4) -- (5,5); 

\draw[thick] (1,4) parabola bend (2.5,3) (4,4); 
\draw[thick] (1,1) parabola bend (2.5,2) (4,1);

\draw[thin] (6,0) -- (6,4); 
\draw[thin] (8,1) -- (8,5); 
\draw[thin] (6,0) -- (8,1); 
\draw[thin] (6,4) -- (8,5); 

\draw[thin] (9,0) -- (9,4); 
\draw[thin] (11,1) -- (11,5); 
\draw[thin] (9,0) -- (11,1); 
\draw[thin] (9,4) -- (11,5);

\draw[thick] (7,4) parabola bend (8.5,2.5) (10,4); 
\draw[thick] (7,1) parabola bend (8.5,2.5) (10,1);

\draw[thin] (12,0) -- (12,4); 
\draw[thin] (14,1) -- (14,5); 
\draw[thin] (12,0) -- (14,1); 
\draw[thin] (12,4) -- (14,5); 

\draw[thin] (15,0) -- (15,4); 
\draw[thin] (17,1) -- (17,5); 
\draw[thin] (15,0) -- (17,1); 
\draw[thin] (15,4) -- (17,5);

\fill[white!30] (13.4,1.6) rectangle (15.6,2.2);

\draw[thick] (13,4) parabola bend (14.2,2.5) (14.2,2.5); 
\draw[thick] (13,1) parabola bend (14.2,2.5) (14.2,2.5); 

\draw[thick] (16,4) parabola bend (14.7,2.5) (14.7,2.5); 
\draw[thick] (16,1) parabola bend (14.7,2.5) (14.7,2.5); 

\node at (14.5,2) {\tiny\bf conical};
\node at (14.5,1.8) {\tiny\bf singularities};

%\draw[thick] (2.5,2.5) ellipse (0.3cm and 0.5cm); 
%\draw[thick] (1.4,2.5) ellipse (0.54cm and 0.9cm); 
%\draw[thick] (3.6,2.5) ellipse (0.54cm and 0.9cm); 

\end{tikzpicture}

\end{center}

\caption{
After the throat connecting two universes is pinched off, conical singularities appear. 
}
\label{bndstt}
\end{figure}

\section{Energy-momentum tensor in a general spherically symmetric but time-dependent spacetime}\label{EMtnsr}

In order to investigate the energy conditions, we need to identify the energy density and the pressures, which are different in the radial direction and the angular or transverse direction, from the geometry. 
In the time-dependent spacetime, even if the spacetime is spherically symmetric, the identifications are not so straightforward. 

The general energy-momentum tensor in the spherically symmetric but time-dependent spacetime has the form in \eqref{EMtensorGeneral}. 
Hereafter, we assume that there is no heat flux, $q_\mu=0$, because we are considering the conserved energy-momentum tensor, 

We assume the dynamical spherically symmetric spacetime in \eqref{ssWHmtrc}. 
In the spacetime \eqref{ssWHmtrc}, the four-velocity vector $u_\mu$ has only time and radial components, $\left(u_\mu\right)=\left( u_t, u_r, 0, 0\right)$ and $u_\mu$ is normalised to be 
\begin{align}
\label{normalisation}
-1=&\, -\e^{-2\nu} {u_t}^2 + \e^{-2\lambda} {u_r}^2 \, .
\end{align}
We also assume $u_t>0$. 
We can also assume the radial vector $s_\mu$ has only time and radial components, $\left(s_\mu\right)=\left( s_t, s_r, 0, 0\right)$ and $s_\mu$ is normalised to be 
\begin{align}
\label{snormalisation}
1=&\, -\e^{-2\nu} {s_t}^2 + \e^{-2\lambda} {s_r}^2 \, .
\end{align}
Because $s_\mu$ is perpendicular to $u_m$, 
\begin{align}
\label{sperpendicular}
0 = s_\mu u^\mu = -\e^{-2\nu} s_t u_t + \e^{-2\lambda} s_r u_r \, ,
\end{align}
we can solve $s_\mu$ with respect to $u_\mu$ as follows, 
\begin{align}
\label{ssolve}
s_t = \pm \e^{\nu - \lambda} u_r \, , \quad s_r = \pm \e^{-\nu + \lambda} u_t \, . 
\end{align}
In the above expressions, the choice of the signature $\pm$ is determined by the double sign. 

Then Eq.~\eqref{EMtensorGeneral} gives 
\begin{align}
\label{EMtensorGeneralComp}
T_{tt} =&\, \left(\rho + p_t \right) {u_t}^2 - p_t \e^{2\nu} + \left( p_r - p_t\right) \e^{2\nu- 2\lambda} {u_r}^2 \, , \quad 
T_{rr} = \left(\rho + p_t \right) {u_r}^2 + p_t \e^{2\lambda} + \left( p_r - p_t\right) \e^{-2\nu+ 2\lambda} {u_t}^2 \, , \nonumber \\
T_{tr} =&\, T_{rt} = \left( \rho + p_r \right) u_r u_t \, , \quad
T_{ij} = p_t \e^{2\xi} {\bar g}_{ij} \, .
\end{align}
Because there are five equations, that is, \eqref{normalisation} and the four equations in \eqref{EMtensorGeneralComp}, we can solve them with respect to the five variables 
$\rho$, $p_r$, $p_t$, $u_t$ and $u_r$ as follows, 
\begin{align}
\label{solutions}
u_t =&\, \frac{1}{\sqrt{2}} \sqrt{ \e^{2\nu} + \sqrt{ \e^{4\nu} + \frac{4\e^{2\nu-2\lambda} {T_{tr}}^2}{\left(\e^{-2\nu} T_{tt} - \e^{-2\lambda} T_{rr} + \e^{-2\xi} {\bar g}^{ij} T_{ij}\right)^2}}} \, , \nonumber \\
u_r =&\, \frac{\e^{\lambda - \nu}}{\sqrt{2}}
\sqrt{ - \e^{2\nu} + \sqrt{ \e^{4\nu} + \frac{4\e^{2\nu-2\lambda} {T_{tr}}^2}{\left(\e^{-2\nu} T_{tt} - \e^{-2\lambda} T_{rr} + \e^{-2\xi} {\bar g}^{ij} T_{ij}\right)^2}}} \, , \nonumber \\
\rho =&\, \e^{-2\nu} T_{tt} - \frac{\e^{-2\nu}}{2} \left( \e^{-2\nu} T_{tt} - \e^{-2\lambda} T_{rr} + \e^{-2\xi} {\bar g}^{ij} T_{ij} \right) 
\left\{ - \e^{2\nu} + \sqrt{ \e^{4\nu} + \frac{4\e^{2\nu-2\lambda} {T_{tr}}^2}{\left(\e^{-2\nu} T_{tt} - \e^{-2\lambda} T_{rr} + \e^{-2\xi} {\bar g}^{ij} T_{ij}\right)^2}} \right\} \nonumber \\
p_r =&\, - \e^{-2\lambda} T_{rr} + \e^{-2\xi} {\bar g}^{ij} T_{ij} \, , \nonumber \\
&\, + \frac{\e^{-2\nu}}{2} \left( \e^{-2\nu} T_{tt} - \e^{-2\lambda} T_{rr} + \e^{-2\xi} {\bar g}^{ij} T_{ij} \right) 
\left\{ - \e^{2\nu} + \sqrt{ \e^{4\nu} + \frac{4\e^{2\nu-2\lambda} {T_{tr}}^2}{\left(\e^{-2\nu} T_{tt} - \e^{-2\lambda} T_{rr} + \e^{-2\xi} {\bar g}^{ij} T_{ij}\right)^2}} \right\} \, , \nonumber \\
p_t =&\, \frac{1}{2} \e^{-2\xi} {\bar g}^{ij} T_{ij} \, . 
\end{align}
By using the Einstein equation $T_{\mu\nu}=\frac{1}{\kappa^2} \left( R_{\mu\nu} - \frac{1}{2} g_{\mu\nu} R \right)$, the energy-momentum tensor can be written in terms of the geometry. 
Then, by using \eqref{solutions}, we find the expressions of $u_t$, $u_r$, $\rho$, $p_r$ and $p_t$ from the geometry, and we may discuss the energy conditions. 

\section{Energy condition}s\label{EC}

Although there is no instability in the two-scalar model \eqref{I8} with \eqref{lambda1} even if the energy conditions are violated, we investigate the energy conditions in this section and consider how the conditions might be satisfied. 

The null energy condition (NEC) and, hence, all the standard energy conditions, including the weak energy condition (WEC), strong energy condition (SEC), and dominant energy condition (DEC), are defined as follows, 
\begin{align}
\label{Eq: energy cond1}
\text{NEC}:& \quad T_{\mu\nu} k^\mu k^\nu \geq 0 \, , \nonumber \\
\text{WEC}:& \quad T_{\mu\nu} V^\mu V^\nu \geq 0 \, ,\nonumber \\
\text{SEC}:& \quad \left(T_{\mu\nu}-\frac{1}{2}g_{\mu\nu}T \right)V^\mu V^\nu \geq 0 \, ,\nonumber \\
\text{DEC}:& \quad T_{\mu\nu} V^\mu V^\nu \geq 0 \, , \ \mbox{and} \ T_{\mu\nu} V^\nu \ \mbox{is not spacelike} \, . 
\end{align}
for any null vector $k^\mu$, $g_{\mu\nu}k^\mu k^\nu =0$, and for any time-like vector $V^\mu$, $g_{\mu\nu}V^\mu V^\nu <0$. 
In terms of $\rho$, $p_r$ and $p_t$, the above expressions in \eqref{Eq: energy cond1} are written as, 
\begin{align}
\label{Eq: energy cond2}
\text{NEC}:& \quad \rho + p_r \geq 0 \, , \quad \rho + p_t \geq 0 \, , \nonumber \\
\text{WEC}:& \quad \rho + p_r \geq 0 \, , \quad \rho + p_t \geq 0 \, , \quad \rho \geq 0 \, ,\nonumber \\
\text{SEC}:& \quad \rho + p_r \geq 0 \, , \quad \rho + p_t \geq 0 \, , \quad \rho +p_r +2 p_t\geq 0 \, ,\nonumber \\
\text{DEC}:& \quad \rho \geq 0 \, , \quad \rho \geq | p_r| \, , \quad \rho \geq | p_t| \, .
\end{align}
Usually, in a wormhole spacetime, some of the energy conditions are always violated, and we consider the energy conditions for the spacetime of the dynamical wormhole in the previous section. 

Before considering the case of \eqref{Hayward} with \eqref{dynmclmdl}, we may consider the simplest wormhole case for comparison, 
\begin{align}
\label{Simplest}
\nu=\lambda=0\, , \quad \e^{2\xi} = r^2 +{r_\mathrm{throat}}^2 \, . 
\end{align}
There is a throat at $r=0$, and the radius of the throat is given by $r_\mathrm{throat}$. 
Then we find 
\begin{align}
\label{curvatureSimplest}
& R_{tt}=0 \, , \quad 
R_{rr} = - \frac{2{r_\mathrm{throat}}^2}{\left( r^2 +{r_\mathrm{throat}}^2 \right)^2} \, , \quad 
R_{tr}=R_{rt} = 0 \, , \nonumber \\
& R_{ij} = \left[ 2 - \frac{r^2}{r^2 + {r_\mathrm{throat}}^2} - \frac{{r_\mathrm{throat}}^2}{r^2 +{r_\mathrm{throat}}^2} \right] \bar{g}_{ij} = \bar{g}_{ij} \, , \quad 
R = \frac{2r^2}{\left( r^2 +{r_\mathrm{throat}}^2 \right)^2} \, .
\end{align}
By using Eq.~\eqref{solutions} and the Einstein equation $T_{\mu\nu}=\frac{1}{\kappa^2} \left( R_{\mu\nu} - \frac{1}{2} g_{\mu\nu} R \right)$, we find that 
the energy density $\rho$, the pressure $p_r$ in the radial direction and the pressure $p_t$ in the angular or transverse direction are given by 
\begin{align}
\label{rhopsSimplest}
\rho =&\, T_{tt} = \frac{1}{\kappa^2} \left( R_{tt} + \frac{1}{2} R \right) 
= \frac{r^2}{\kappa^2\left( r^2 +{r_\mathrm{throat}}^2 \right)^2} \, , \nonumber \\
p_r =&\, - T_{rr} + \frac{1}{r^2 +{r_\mathrm{throat}}^2} \bar{g}^{ij} T_{ij} 
= \frac{ 4{r_\mathrm{throat}}^2 - r^2}{\kappa^2 \left( r^2 +{r_\mathrm{throat}}^2 \right)^2} \, , \nonumber \\
p_t =&\, \frac{1}{2} \e^{-2\xi} {\bar g}^{ij} T_{ij} =0 \, .
\end{align}
At the throat $r=0$, we obtain
\begin{align}
\label{rhopsSimplestthroat}
\rho=0\, , \quad p_r = \frac{4}{\kappa^2 {r_\mathrm{throat}}^2}\, , \quad p_t=0 \, .
\end{align}
Therefore, NEC, WEC and SEC are satisfied, but DEC is violated. 

Because, in the case of \eqref{Hayward} with \eqref{dynmclmdl}, the expressions of the curvatures become very complicated, we may only give the expressions at the throat $r=0$, 
\begin{align}
\label{curvaturethroat}
R_{tt}=&\, - \frac{2}{\gamma^2} - \frac{2{\ddot\Xi}_{02}(t)}{r_\mathrm{throat} + \Xi_{02}(t)} \, , \quad 
R_{rr}= \frac{1}{\gamma^2} - \frac{2}{r_\mathrm{throat} \left( r_\mathrm{throat} + \Xi_{02}(t)\right)} \, , \quad 
R_{tr}=R_{rt} = 0\, , \nonumber \\
R_{ij} =&\, \left[ 2 + {\dot\Xi}_{02}(t)^2 + {\ddot\Xi}_{02}(t)\left(r_\mathrm{throat} + \Xi_{02}(t)\right) - \frac{r_\mathrm{throat} + \Xi_{02}(t)}{r_\mathrm{throat}} \right] \bar{g}_{ij} \, , \nonumber \\
R =&\, \frac{2}{\gamma^2} + \frac{4{\ddot\Xi}_{02}(t)}{r_\mathrm{throat} + \Xi_{02}(t)} - \frac{4}{r_\mathrm{throat} \left( r_\mathrm{throat} + \Xi_{02}(t)\right)} 
+ \frac{2}{\left( r_\mathrm{throat} + \Xi_{02}(t)\right)^2} \left( 2 + {\dot\Xi}_{02}(t)^2 \right) \, .
\end{align}
Then we find the expressions of $\rho$, $p_r$ and $p_t$ at the throat
\begin{align}
\label{rhopsthroat}
\rho =&\, \frac{1}{\kappa^2} \left\{ - \frac{1}{\gamma^2} 
 - \frac{2}{r_\mathrm{throat} \left( r_\mathrm{throat} + \Xi_{02}(t)\right)} 
+ \frac{2 + {\dot\Xi}_{02}(t)^2}{\left( r_\mathrm{throat} + \Xi_{02}(t)\right)^2} \right\} \, , \nonumber \\
p_r =&\, \frac{1}{\kappa^2} \left\{ - \frac{2}{\gamma^2} + \frac{4}{r_\mathrm{throat} \left( r_\mathrm{throat} + \Xi_{02}(t)\right)} 
+ \frac{{\dot\Xi}_{02}(t)^2}{\left( r_\mathrm{throat} + \Xi_{02}(t)\right)^2} + \frac{2\Xi_{02}(t)}{r_\mathrm{throat}\left( r_\mathrm{throat} + \Xi_{02}(t)\right)^2} \right\} \, , \nonumber \\
p_t =&\, \frac{1}{\kappa^2} \left\{ 
 - \frac{1}{\gamma^2} - \frac{{\ddot\Xi}_{02}(t)}{r_\mathrm{throat} + \Xi_{02}(t)} + \frac{1}{r_\mathrm{throat} \left( r_\mathrm{throat} + \Xi_{02}(t)\right)} \right\} \, .
\end{align}
When $\gamma\to \infty$ and $\Xi_{02}(t)=0$, the above expressions in \eqref{rhopsthroat} reduce to those in \eqref{rhopsSimplest}. 

In the static case that $\gamma$ is finite but $\Xi_{02}(t)=0$, we find 
\begin{align}
\label{rhopsthroatst}
\rho = - \frac{1}{\kappa^2\gamma^2} \, , \quad 
p_r = \frac{1}{\kappa^2} \left\{ - \frac{2}{\gamma^2} + \frac{4}{{r_\mathrm{throat}}^2} \right\} \, , \quad 
p_t = \frac{1}{\kappa^2} \left\{ - \frac{1}{\gamma^2} + \frac{1}{{r_\mathrm{throat}}^2} \right\} \, ,
\end{align}
and therefore we obtain, 
\begin{align}
\label{rhopsthroatstEC}
\rho<0\, , \quad \rho + p_r =&\, \frac{1}{\kappa^2} \left\{ - \frac{3}{\gamma^2} + \frac{4}{{r_\mathrm{throat}}^2} \right\} \, , \quad 
\rho + p_t = \frac{1}{\kappa^2} \left\{ - \frac{2}{\gamma^2} + \frac{1}{{r_\mathrm{throat}}^2} \right\} \, , \nonumber \\
\rho + p_r + 2p_t =&\, \frac{1}{\kappa^2} \left\{ - \frac{5}{\gamma^2} + \frac{6}{{r_\mathrm{throat}}^2} \right\} \, .
\end{align}
Because the energy density is negative, WEC and DEC are violated. 
On the other hand, if $\gamma^2 > 2 {r_\mathrm{throat}}^2$, NEC and SEC are not broken. 

In the dynamical case that $\gamma$ is finite and $\Xi_{02}(t)$ is a non-trivial function, we have 
\begin{align}
\label{rhopsthroatgen}
\rho =&\, \frac{1}{\kappa^2} \left\{ - \frac{1}{\gamma^2} 
 - \frac{2}{r_\mathrm{throat} \left( r_\mathrm{throat} + \Xi_{02}(t)\right)} 
+ \frac{2 + {\dot\Xi}_{02}(t)^2}{\left( r_\mathrm{throat} + \Xi_{02}(t)\right)^2} \right\} \, , \nonumber \\
\rho + p_r =&\, \frac{1}{\kappa^2} \left\{ - \frac{3}{\gamma^2} + \frac{2}{r_\mathrm{throat} \left( r_\mathrm{throat} + \Xi_{02}(t)\right)} 
+ \frac{2 + 2 {\dot\Xi}_{02}(t)^2}{\left( r_\mathrm{throat} + \Xi_{02}(t)\right)^2} + \frac{2\Xi_{02}(t)}{r_\mathrm{throat}\left( r_\mathrm{throat} + \Xi_{02}(t)\right)^2} \right\} \, , \nonumber \\
\rho + p_t =&\, \frac{1}{\kappa^2} \left\{ 
 - \frac{2}{\gamma^2} - \frac{{\ddot\Xi}_{02}(t)}{r_\mathrm{throat} + \Xi_{02}(t)} - \frac{1}{r_\mathrm{throat} \left( r_\mathrm{throat} + \Xi_{02}(t)\right)} 
+ \frac{2 + {\dot\Xi}_{02}(t)^2}{\left( r_\mathrm{throat} + \Xi_{02}(t)\right)^2} \right\} \, , \nonumber \\
\rho + p_r + 2p_t =&\, \frac{1}{\kappa^2} \left\{ 
 - \frac{5}{\gamma^2} - \frac{2{\ddot\Xi}_{02}(t)}{r_\mathrm{throat} + \Xi_{02}(t)} + \frac{4}{r_\mathrm{throat} \left( r_\mathrm{throat} + \Xi_{02}(t)\right)} 
+ \frac{2 + 2 {\dot\Xi}_{02}(t)^2}{\left( r_\mathrm{throat} + \Xi_{02}(t)\right)^2} \right. \nonumber \\
& \qquad \left. + \frac{2\Xi_{02}(t)}{r_\mathrm{throat}\left( r_\mathrm{throat} + \Xi_{02}(t)\right)^2} 
\right\} \, .
\end{align}
The above expressions suggest that the energy conditions might be satisfied if we include the contribution of $\Xi_{02}(t)$. 
If we choose ($\alpha>0$ and $r_0>0$)
\begin{align}
\label{choice}
\ln \left( r_\mathrm{throat} + \Xi_{02}(t)\right) = - \frac{\alpha}{2} t^2 \, , \quad r_\mathrm{throat} + \Xi_{02}(t) = r_0 \e^{-\frac{\alpha}{2} t^2}\, , 
\end{align}
with positive constants $\alpha$ and $r_0$, becasue we find 
\begin{align}
\label{choice2}
\frac{d\ln \left( r_\mathrm{throat} + \Xi_{02}(t)\right)}{dt} =&\, \frac{\dot\Xi_{02}(t)}{r_\mathrm{throat} + \Xi_{02}(t)} = - \alpha t \, , \nonumber \\
\frac{d^2\ln \left( r_\mathrm{throat} + \Xi_{02}(t)\right)}{dt^2}
=&\, \frac{\ddot\Xi_{02}(t)}{r_\mathrm{throat} + \Xi_{02}(t)} - \frac{\dot\Xi_{02}(t)^2}{\left(r_\mathrm{throat} + \Xi_{02}(t)\right)^2} =- \alpha \, .
\end{align}
When $|t|\to \infty$, we find 
\begin{align}
\label{choice3}
\rho\to&\, \frac{\alpha^2 t^2}{\kappa^2} \, , \quad 
p_r \to \frac{\alpha^2 t^2}{\kappa^2} \, , \quad 
p_t \to - \frac{\alpha^2 t^2}{\kappa^2} \, , \quad 
\rho + p_r \to \frac{2\alpha^2 t^2}{\kappa^2} \, , \nonumber \\
\rho + p_t \to&\, \frac{1}{\kappa^2} \left\{ - \frac{2}{\gamma^2} + \alpha + \frac{1}{{r_\mathrm{throat}}^2} \right\} \, , \quad 
\rho + p_r + 2p_t \to \frac{1}{\kappa^2} \left\{ 
 - \frac{5}{\gamma^2} + 2\alpha + \frac{6}{{r_\mathrm{throat}}^2} \right\} \, , 
\end{align}
Then DEC is not violated, and if we choose 
\begin{align}
\label{gg}
\alpha> \frac{2}{\gamma^2} - \frac{1}{{r_\mathrm{throat}}^2} \, , \ \frac{5}{2\gamma^2} - \frac{3}{{r_\mathrm{throat}}^2} \, ,
\end{align}
NEC, WEC and SEC are not violated. 

Then we have shown that the energy conditions might be recovered in the dynamical spacetime as found in \cite{Katsuragawa:2025zcy}. 

\section{Simpson-Visser black bounce}\label{SVbb}

We may consider the time-dependent extension of the Simpson-Visser black bounce~\cite{Simpson:2018tsi}. 
The metric of the original Simpson-Visser black bounce is given by 
\begin{align}
\label{SVmetric}
\e^{2\nu}=\e^{-2\lambda}= 1 - \frac{2M}{\sqrt{r^2 + {r_\mathrm{throat}}^2}}\, , \quad 
\e^{2\xi} = r^2 + {r_\mathrm{throat}}^2\, ,
\end{align}
in the notation of \eqref{ssWHmtrc}. 
In Eq.~\eqref{SVmetric}, the radial coorsinate $r$ runs from $-\infty$ to $+\infty$. 
When $r_\mathrm{throat}=0$, the spacetime reduces to the Schwarzschild metric. 
On the other hand, the spacetime with $r_\mathrm{throat}\neq 0$ has no curvature singularity and therefore the spacetime Eq.~\eqref{SVmetric} can be regarded as a regularisation of the Schwarzschild black hole. 
When $r_\mathrm{throat}>2M$, the horizon does not appear, and the spacetime becomes a wormhole geometry where a universe corresponding to the positive $r$ is connected with another universe corresponding to the negative $r$. 
On the other hand, when $r_\mathrm{throat}<2M$, the geometry has a throat inside the black hole and two universes corresponding to positive and negative $r$, respectively, are connected. 
Because the throat is time-like, not as in the standard wormhole, where the throat is space-like, the throat is one-way; that is, any object going through the throat from one universe to another universe cannot go back to the universe where the object was. 
The geometry of the throat is similar to that of the cosmological big bounce, and therefore the geometry is called ``black bounce''. 
The case $r_\mathrm{throat}=2M$ is extremal, and the Penrose diagram of the spacetime is shown in \cite{Simpson:2018tsi}.

The arguments around Eq.~\eqref{r4} tell that $\e^{2\nu}$ and $\e^{2\lambda}$ cannot be time-dependent but only $\e^{2\xi}$ can depend on time when $r_\mathrm{throat}<2M$. 
The time dependence of $\e^{2\xi}$ is, as in \eqref{xi1}, given by 
\begin{align}
\label{SVxi1}
\e^{\xi\left(t,r\right)} = \Xi_1 (r) + \e^{2\nu(r)} \Xi_2(r,t)= \Xi_1 (r) + \left( 1 - \frac{2M}{\sqrt{r^2 + {r_\mathrm{throat}}^2}} \right) \Xi_2(r,t)
\, , \quad \Xi_2(r,t) \equiv \int^t dt_1 \int^{t_1} dt_2 \Xi \left(t_2, r\right) \, .
\end{align}
As in \eqref{xi1}, $\Xi_1 (r)$ can be an arbitrary function of $r$ and $\Xi_2(r,t)$ can also be an arbitrary function of $t$ and $r$ as long as a smooth function. 

When $r_\mathrm{throat}>2M$, we may also consider the model similar to \eqref{dynmclmdl}
\begin{align}
\label{SVdynmclmdl}
\Xi_1 (r) = \sqrt{r^2 +{r_\mathrm{throat}}^2}\, , \quad \Xi_2=\left( 1 - \frac{2M}{\sqrt{r^2 + {r_\mathrm{throat}}^2}} \right)^{-1} \Xi_{02}(t) \quad \mbox{that is} \quad 
\e^{\xi} = \sqrt{r^2 +{r_\mathrm{throat}}^2} + \Xi_{02}(t) \, .
\end{align}
In the model~\eqref{SVdynmclmdl}, the throat radius ${\bar r}_\mathrm{throat}$ is time-dependent, again 
\begin{align}
\label{SVthdynml}
{\bar r}_\mathrm{throat} = r_\mathrm{throat} + \Xi_{02}(t) \, .
\end{align}
When the throat radius ${\bar r}_\mathrm{throat}$ becomes very large, the universe might be swallowed by the wormhole. 
We should note that we cannot construct a model where there is a transition between the standard wormhole and the black bounce because $\e^{2\nu}$ and $\e^{2\lambda}$ cannot depend on time. 

When $r$ is small, we find 
\begin{align}
\label{SVorigin}
\e^{2\nu} = \e^{-2\lambda} = 1 - \frac{2M}{r_\mathrm{throat}} + \frac{Mr^2}{{r_\mathrm{throat}}^3} + \mathcal{O}\left( r^4 \right)\, , 
\end{align}
and therefore at the throat $r=0$, we obtain, 
\begin{align}
\label{curvaturesWHcheckorigin2}
R_{tt}=& \left( 1 - \frac{2M}{r_\mathrm{throat}} \right)^2 \frac{2M}{{r_\mathrm{throat}}^3} - \frac{2{\ddot\Xi}_{02}(t)}{r_\mathrm{throat} + \Xi_{02}(t)} \, ,\quad 
R_{rr} = - \frac{2M}{{r_\mathrm{throat}}^3} - \frac{2}{r_\mathrm{throat} \left( r_\mathrm{throat} + \Xi_{02}(t)\right)} \, ,\quad 
R_{tr} = R_{rt} = 0 \, , \nonumber \\
R_{ij} =&\, \left[ 2 + \left( 1 - \frac{2M}{r_\mathrm{throat}} \right)^{-1} \left( {\dot\Xi}_{02}(t)^2 + {\ddot\Xi}_{02}(t) \left( r_\mathrm{throat} + \Xi_{02}(t) \right) \right)
 - \left( 1 - \frac{2M}{r_\mathrm{throat}} \right) \frac{ r_\mathrm{throat} + \Xi_{02}(t) }{r_\mathrm{throat}} \right] \bar{g}_{ij}\, , \nonumber \\
R=&\, - 2 \left( 1 - \frac{2M}{r_\mathrm{throat}} \right) \frac{2M}{{r_\mathrm{throat}}^3} + 4 \left( 1 - \frac{2M}{r_\mathrm{throat}} \right)^{-1} \frac{{\ddot\Xi}_{02}(t)}{r_\mathrm{throat} + \Xi_{02}(t)} \nonumber \\
&\, - 4 \left( 1 - \frac{2M}{r_\mathrm{throat}} \right) \frac{1}{r_\mathrm{throat} \left( r_\mathrm{throat} + \Xi_{02}(t)\right)} 
+ \frac{4}{\left( r_\mathrm{throat} + \Xi_{02}(t) \right)^2} + 2 \left( 1 - \frac{2M}{r_\mathrm{throat}} \right)^{-1} \frac{{\dot\Xi}_{02}(t)^2}{\left(r_\mathrm{throat} + \Xi_{02}(t)\right)^2} \, ,
\end{align}
which gives 
\begin{align}
\label{solveSV}
\rho =&\, \frac{1}{\kappa^2} \left\{
 - 2 \left( 1 - \frac{2M}{r_\mathrm{throat}} \right) \frac{1}{r_\mathrm{throat} \left( r_\mathrm{throat} + \Xi_{02}(t)\right)} 
+ \frac{2}{\left( r_\mathrm{throat} + \Xi_{02}(t) \right)^2} 
+ \left( 1 - \frac{2M}{r_\mathrm{throat}} \right)^{-1} \frac{{\dot\Xi}_{02}(t)^2}{\left(r_\mathrm{throat} + \Xi_{02}(t)\right)^2} 
\right\} \, , \nonumber \\
p_r =&\, \frac{1}{\kappa^2\left(r_\mathrm{throat} + \Xi_{02}(t)\right)^2} 
\left\{ 6 + \left( 1 - \frac{2M}{r_\mathrm{throat}} \right)^{-1} \left( 3 {\dot\Xi}_{02}(t)^2 + 4{\ddot\Xi}_{02}(t) \left( r_\mathrm{throat} + \Xi_{02}(t) \right) \right) 
 - 2 \left( 1 - \frac{2M}{r_\mathrm{throat}} \right) \frac{ r_\mathrm{throat} + \Xi_{02}(t) }{r_\mathrm{throat}} 
\right\}\, , \nonumber \\
p_t =&\, \frac{1}{\kappa^2} \left[ 
\left( 1 - \frac{2M}{r_\mathrm{throat}} \right) \frac{2M}{{r_\mathrm{throat}}^3} 
 - \left( 1 - \frac{2M}{r_\mathrm{throat}} \right)^{-1} \frac{{\ddot\Xi}_{02}(t)}{r_\mathrm{throat} + \Xi_{02}(t)} 
+ \left( 1 - \frac{2M}{r_\mathrm{throat}} \right) \frac{1}{r_\mathrm{throat} \left( r_\mathrm{throat} + \Xi_{02}(t)\right)} 
\right] \, . \nonumber \\
\end{align}
Then, as in the last section, by adjusting the function $\Xi_{02}(t)$, some of the energy conditions could be satisfied. 

\section{Thermodynamics of dynamical black bounce spacetime}\label{TDdbbs}

In this section, we consider the thermodynamics of the time-dependent extension of the Simpson-Visser black bounce in the last section. 
In the time-dependent background, there is no time-like Killing vector, and therefore, we cannot define the Hawking temperature by using the surface gravity. 
Furthermore, if we calculate the radiation in a time-dependent background, the distribution of the radiation does not coincide with the thermal Boltzmann distribution, which makes it impossible for us to define the temperature. 
We can use, however, the adiabatic approximation as long as the time development is very slow compared with the Planck scale. 
The adiabatic condition could be given by 
\begin{align}
\label{adabatic}
\left| \frac{\dot \Xi_{02}(t)}{\Xi_{02}(t)} \right| \ll \frac{1}{\kappa}\, ,
\end{align}
which is usually satisfied. 

Because we consider the model with horizons, instead of \eqref{SVdynmclmdl}, we choose 
\begin{align}
\label{SVdynmclmdlhrzn}
\Xi_1 (r) = \sqrt{r^2 +{r_\mathrm{throat}}^2}\, , \quad \Xi_2= \Xi_{02}(t) \, ,
\end{align}
that is 
\begin{align}
\label{SVmetricTD}
\e^{2\nu}=\e^{-2\lambda}= 1 - \frac{2M}{\sqrt{r^2 + {r_\mathrm{throat}}^2}}\, , \quad 
\e^{\xi} = \sqrt{r^2 +{r_\mathrm{throat}}^2} + \left( 1 - \frac{2M}{\sqrt{r^2 + {r_\mathrm{throat}}^2}} \right) \Xi_{02}(t) \, .
\end{align}
We now investigate the thermodynamics of the model \eqref{SVmetricTD}. 

The surface gravity $\tilde\kappa$ in the adiabatic approximation is given by 
\begin{align}
\label{srfcgr}
\tilde\kappa = \frac{1}{2} \left. \frac{d\e^{2\nu}}{d\e^\xi} \right|_{\sqrt{r^2 + {r_\mathrm{throat}}^2}=2M} \frac{1}{2} \left. \frac{\frac{d\e^{2\nu}}{dr}}{{d\e^\xi}{dr}} \right|_{r= \sqrt{4M^2 - {r_\mathrm{throat}}^2}}
= \left. \frac{\frac{Mr}{\left( r^2 + {r_\mathrm{throat}}^2 \right)^\frac{3}{2}}}{\frac{r}{\sqrt{r^2 +{r_\mathrm{throat}}^2}}} \right|_{r= \sqrt{4M^2 - {r_\mathrm{throat}}^2}} 
= \frac{1}{4M} \, ,
\end{align}
which gives the standard Hawking temperature $T_\mathrm{H}$, 
\begin{align}
\label{TH}
T_\mathrm{H} = \frac{\tilde\kappa}{2\pi} = \frac{1}{8\pi M}\, .
\end{align}
Therefore, the Hawking temperature is identical to that of the Schwarzschild black hole. 

We now consider the ADM mass, which is often identified with the thermodynamical energy. 
Because the areal radius is $\e^\xi$, when $\e^\xi$ is large, we find 
\begin{align}
\label{ADMm}
\e^{2\nu}= 1 - \frac{2M}{\e^\xi - \left( 1 - \frac{2M}{\sqrt{r^2 + {r_\mathrm{throat}}^2}} \right) \Xi_{02}(t)} = 1 - 2M \e^{-\xi} + \mathcal{O}\left( \e^{-2\xi} \right) \, .
\end{align}
Therefore, the ADM mass is given by $M$, which is also identical to that in the Schwarzschild black hole. 

We may define the entropy $\mathcal{S}$ by using the surface area of the horizon, as the Bekenstein-Hawking entropy~\cite{Bekenstein:1972tm, Bekenstein:1973ur, Hawking:1975vcx}
\begin{align}
\label{entrpy}
\mathcal{S}_\mathrm{BH} = \left. \frac{4\pi \e^{2\xi}}{4} \right|_{\sqrt{r^2 + {r_\mathrm{throat}}^2}=2M\ \mathrm{that\ is}\ \e^\xi = 2M}
= \pi \left( 2M \right)^2 \, .
\end{align}
The expression is also identical with that of the Schwarzschild black hole, including the static case $\Xi_{02}(t)=0$. 

The above results indicate that the thermodynamics of the Simpson-Visser black bounce and its time-dependent variation do not differ from those of the Schwarzschild black hole. 

\section{Shadows of non-stationary wormholes}\label{shadows-non-stationary-wormholes}

As a final section, we use the so-called ray-tracing method \cite{Gralla:2019xty} to study the optical appearance of some of the examples of dynamical spacetimes previously considered, namely the Simpson-Visser and Hayward wormholes.
All of these examples are particular cases of the more general metric~\eqref{ssWHmtrc}, and they are described by line elements of the form
\begin{align}
\label{eq:metric_ray_tracing}
ds^2 = - \e^{2\nu (r)} dt^2 + \e^{-2\nu (r)} dr^2 + \e^{2\xi (t,r)} \left( d\vartheta^2 + \sin^2\vartheta \, d\varphi^2 \right)\, .
\end{align}
The areal radius function $\e^\xi$ encodes the time-dependence, and either takes the form
\begin{align}
\label{eq:areal_case_1}
\e^\xi = \sqrt{r^2+{\rt}^2}+\Xi_{02}(t)
\end{align}
or, alternatively,
\begin{align}
\label{eq:areal_case_2}
\e^\xi = \sqrt{r^2+{\rt}^2}+\e^{2\nu}\Xi_{02}(t)\, .
\end{align}

We now write the equations of motion for null geodesics.
Due to spherical symmetry, we may restrict ourselves to the equatorial plane $\vartheta=\pi/2$ with no loss of generality.
Then, the equations for coordinates $t$ and $\varphi$ read
\begin{align}
\frac{d}{d\sigma}\left(\e^{2\nu} k^t \right) + \frac{(k^\varphi)^2}{2}\frac{\partial}{\partial t}\left( \e^{2\xi} \right) = 0\, ,
\quad
\frac{d}{d\sigma}\left( \e^{2\xi}k^\varphi \right) = 0\, ,
\end{align}
where $\sigma$ is an affine parameter along the null geodesic $x^\mu(\sigma)$ with $k^\mu = dx^\mu/d\sigma$ being its tangent null vector.
We further consider the adiabatic approximation~\eqref{adabatic} such that time changes in the geometry can be neglected when light rays travel along geodesics.
In this regime, the partial derivative term in the left equation vanishes, and we get the two conserved quantities
\begin{align}
\e^{2\nu} k^t = E\, ,
\quad
\e^{2\xi} k^\varphi = L\, .
\end{align}
Thus, in what follows, we shall treat $\Xi_{02}(t)\equiv\Xi$ as an additional parameter on which metric functions may depend.
The motion is then described by the first integral
\begin{align}
\left(\frac{d r}{d\sigma}\right)^2 = \frac{1}{b^2} - V(r)\, ,
\end{align}
where we have re-parametrized the affine parameter $\sigma\rightarrow L\sigma$ to absorb the $L^2$ factor.
This equation only depends on the so-called impact parameter $b=L/E$ and the effective potential $V(r)=e^{2\nu-2\xi}$.

The equation for the trajectory of null geodesics reads
\begin{align}
\frac{d\varphi}{dr} = \pm \frac{b\ \e^{-\xi}}{\sqrt{\e^{2\xi}-b^2 \e^{2\nu}}} \, ,
\label{eq:trajectory}
\end{align}
where the plus (minus) sign corresponds to outgoing (incoming) light rays.

We model the accretion disk as being geometrically and optically thin and lying on the equatorial plane.
We further assume that the emission is isotropic with specific intensity $I^\mathrm{em}_\omega$ in its rest frame; and that it is static, i.e., the redshift is only due to gravitational redshift.
Since $I^\mathrm{em}_\omega/\omega^3$ is conserved along any given geodesic, the observed specific intensity is given by $I^\mathrm{ob}_{\omega'} = (\omega' /\omega)^3 I^\mathrm{em}_\omega = \e^{3\nu}I^\mathrm{em}_\omega$. 
The integrated observed intensity $I^\mathrm{ob} = \int I^\mathrm{ob}_{\omega'} d\omega'$ is then given by the sum of all contributions for $m$ intersections with the accretion disk
\begin{align}
I^\mathrm{ob}(b) = \sum_{m=1}^\infty \left. \e^{4\nu(r)}I^\mathrm{em}(r)\right|_{r=r_m(b)}\, .
\label{eq:observed-intensity}
\end{align}
The functions $r_m(b)$ are dubbed \textit{transfer functions}, and give the radial coordinate $r$ of the $m$-th intersection with the accretion disk for a geodesic with impact parameter $b$.
Despite intersecting an arbitrarily large number of times, light rays with large $m$ contribute negligibly to the total intensity \cite{Gralla:2019xty, Bisnovatyi-Kogan:2022ujt}.
Therefore, we only consider up to the first three orders of $m$, which we dub as direct ($m=1$), lensed ($m=2$) and photon ring ($m=3$) emission.

To model the emission, we will use the parametrised family of SU models~\cite{Gralla:2020srx, Paugnat:2022qzy, Cardenas-Avendano:2022csp}, whose expression is given by
\begin{align}
 I^\mathrm{em}(r;\mu,\sigma,\beta) = \frac{\exp \left( -\frac{1}{2}\left(\beta+\arcsinh\left(\frac{r-\mu}{\sigma}\right)\right)^2 \right)}{ \sqrt{(r-\mu)^2+\sigma^2} }\, .
\label{eq:su-model}
\end{align}
In these emission profiles, $\mu$, $\sigma$ and $\beta$ parametrise the location of the emission peak, its width and the asymmetry of the distribution, respectively.
In all the considered examples, we will assume that $\mu$ coincides with the inner edge $\ri$ of the accretion disk, which is placed at the (outer) event horizon for a black hole or at the throat for a horizonless wormhole.
For an SV black hole, the event horizon is given by $\rhor=\sqrt{4M^2-{\rt}^2}$ if $\rt\leq2M$, while the throat is at $r=0$.
On the other hand, a Hayward black hole has its outer event horizon at~\eqref{rpm} with the plus sign if condition~\eqref{condition2} is verified, whereas the throat of a Hayward wormhole is also at $r=0$.

\begin{figure}[ht!]
\includegraphics{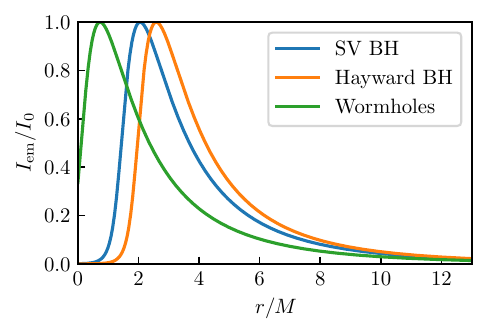}
\caption{
SU emission profiles~\eqref{eq:su-model} used for each example, normalised with respect to the maximum emitted intensity $I_0$.
For a SV black hole with $\rt=1.5M$ (blue) the value of $\mu$ is set at $\ri=\sqrt{7}/2M\approx1.32M$, for Hayward black hole with $\gamma=0.5M$ (orange) we have $\ri\approx1.85M$, whilst for both SV and Hayward wormholes (green) $\mu=\ri=0$.
The remaining parameters have been set to $\sigma=0.5M$ and $\beta=-2$ in all cases. 
}
\label{fig:emission-models}
\end{figure}

%%%%%%%%%%%%%%%%%%%%%%

\subsection{Shadows of dynamical Simpson-Visser spacetime}\label{shadows-SV}

\begin{figure}[ht!]
\includegraphics[width=\textwidth]{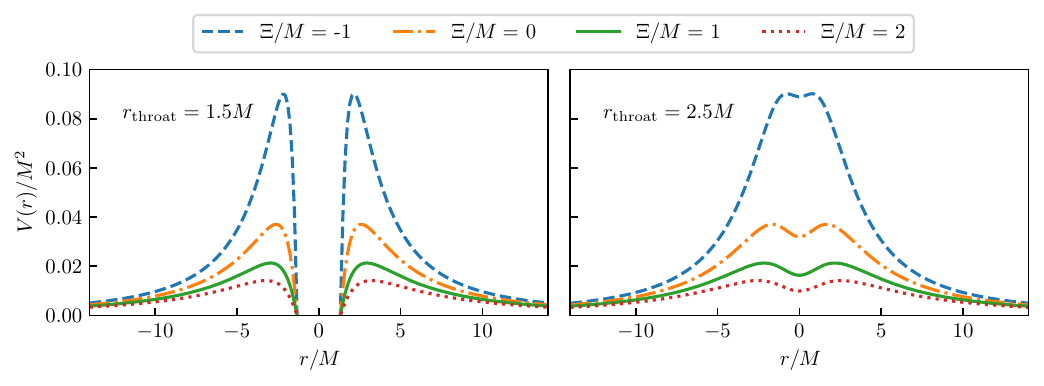}
\caption{Simpson-Visser potential for $\rt=1.5M$ (left) and $\rt=2.5M$ (right) for different values of the parameter $\Xi$.
The left panel corresponds to a black hole, whereas the right panel corresponds to a traversable wormhole.}
\label{fig:SV_potential}
\end{figure}

We now study the Simpson-Visser black bounce with $\e^{2\nu}$ given by Eq.~\eqref{SVmetric} and $\e^{2\xi}$ in Eq.~\eqref{eq:areal_case_1}.
The optical appearance of this black bounce has already been studied in \cite{Guerrero:2021ues}, so it gives a good basis for investigating the effect of an adiabatic temporal evolution on its shadow.
The explicit expression for the potential is
\begin{align}
V(r) = \left( \sqrt{r^2+{\rt}^2}+\Xi \right)^{-2}\left( 1-\frac{2M}{\sqrt{r^2+{\rt}^2}} \right)\, .
\end{align}
Just like the static case \cite{Simpson:2018tsi}, the potential only depends on $\rt$ through the radial function $R(r)=\sqrt{r^2+{\rt}^2}$, so the value of the critical impact parameter $b_c$ will not depend on $\rt$.
The shape of this potential is depicted in FIG.~\ref{fig:SV_potential} for two qualitatively different geometries (black hole with $\rt=1.5M$, traversable wormhole with $\rt=2.5M$) and different choices for $\Xi$.

The critical points of the potential satisfy the following condition
\begin{align}
\label{eq:critical_points}
r\left( \frac{9}{2} + \frac{\Xi}{M} \pm 3\sqrt{\frac{9}{4}+\frac{\Xi}{M}} - \frac{R(r)^2}{M^2} \right) = 0\, .
\end{align}
Likewise, in the static case, we have a local minimum at $r=0$ associated with stable circular orbits\footnote{These stable circular orbits allow for trapped long-living modes which could potentially destabilise the system.} or an anti-photon sphere.
Moreover, we have two values
\begin{align}
\frac{R_\pm^2}{M^2} = \frac{9}{2} + \frac{\Xi}{M} \pm 3\sqrt{\frac{9}{4}+\frac{\Xi}{M}}
\end{align}
for which the condition~\eqref{eq:critical_points} holds: the one with a plus sign corresponds to a maximum, whereas the minus sign corresponds to a minimum.
These extremal points exist for $\Xi/M\geq-9/4$, and they degenerate when the inequality is saturated.
However, we have only circular null orbits for $R_+$ with $\Xi/M > -2$ since the potential is non-positive at $R_-$ for $-9/4 \geq \Xi/M \geq -2$ as well as at $R_-$ for all values of $\Xi$.

The photon sphere is placed at coordinate $\rp = \sqrt{R_+^2-{\rt}^2}$ and the critical impact parameter $b_c$ is found at
\begin{align}
\label{eq:b_crit}
b_c=\frac{1}{\sqrt{V(\rp)}} = \frac{M}{2}\sqrt{ 18\left( 3+\sqrt{9+\frac{4\Xi}{M}} \right) +\frac{4\Xi}{M}\left( 9+\frac{\Xi}{M}+2\sqrt{9+\frac{4\Xi}{M}} \right) } \, . 
\end{align}

\begin{figure}[ht!]
\centering
% SV black hole
\begin{subfigure}[c]{0.32\textwidth} % Left
\centering
\caption{$\Xi/M = -1$}
\includegraphics[width=\textwidth]{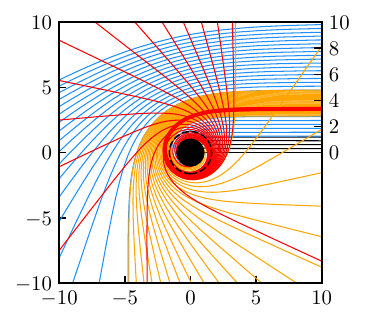}
\end{subfigure}
\begin{subfigure}[c]{0.32\textwidth} % Middle
\centering
\caption{$\Xi/M = 0$}
\includegraphics[width=\textwidth]{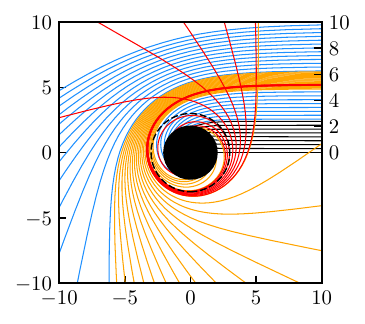}
\end{subfigure}
\begin{subfigure}[c]{0.32\textwidth} % Right
\centering
\caption{$\Xi/M = 1$}
\includegraphics[width=\textwidth]{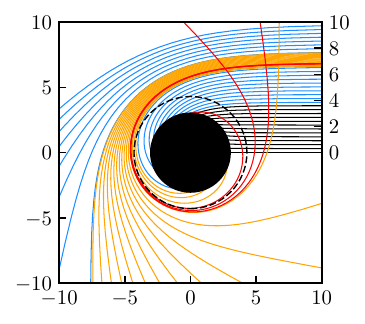}
\end{subfigure}

% SV wormhole
\begin{subfigure}[c]{0.32\textwidth} % Left
\centering
\includegraphics[width=\textwidth]{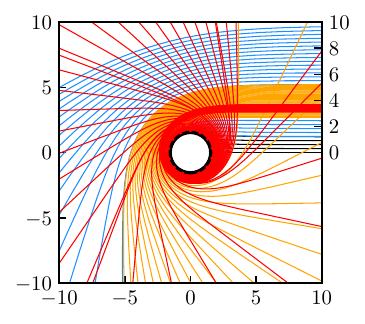}
\end{subfigure}
\begin{subfigure}[c]{0.32\textwidth} % Middle
\centering
\includegraphics[width=\textwidth]{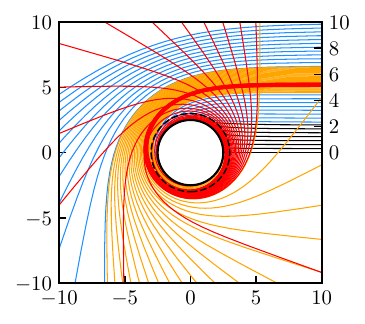}
\end{subfigure}
\begin{subfigure}[c]{0.32\textwidth} % Right
\centering
\includegraphics[width=\textwidth]{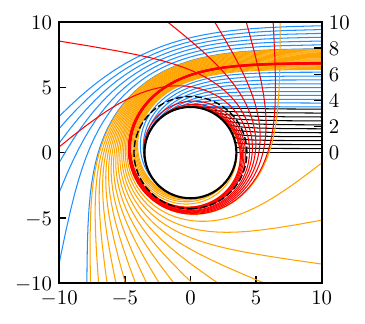}
\end{subfigure}
 
\caption{
Ray-tracing for an SV black hole with $\rt=1.5$ (top row) and a traversable wormhole with $\rt=2.5$ (bottom row) for different values of $\Xi$ in each column.
They have been plotted in terms of the areal radius $\e^\xi$ and in units of $M=1$.
Light trajectories are coloured in blue, orange and red if they correspond to direct, lensed or photon ring images, respectively; while those coloured in black are those that do not hit the disk and make up the inner shadow.
In all cases, the dashed black circle represents the photon sphere $\rp$, while the central disk corresponds to the region inside the inner edge of the disk, filled with black if it is at the event horizon $r_\mathrm{hor}$ or with white if it is at the $\rt$.
One effect is the growth of both the photon ring and critical impact parameter as $\Xi$ increases.
Moreover, since the spacings between impact parameters are the same for all figures, we see that when $\Xi$ grows, the lensed and photon rings become narrower as there are fewer trajectories.
}
\label{fig:SV-ray-tracing}
\end{figure}

Integrating the trajectory equation~\eqref{eq:trajectory}, we obtain points $(r,\phi)$ of the trajectory from which we may plot $(\e^\xi,\phi)$ in Cartesian polar coordinates as in FIG.~\ref{fig:SV-ray-tracing}. 

%%%%%%% RAY-TRACING %%%%%%%
%%%%%%%%%%%%%%%%%%%%%%%%%%%%

%%%% TRANSFER FUNCTIONS %%%%
\begin{figure}[ht!]
\centering
% SV black hole
\begin{subfigure}[c]{0.32\textwidth} % Left
\centering
\caption{$\Xi/M = -1$}
\includegraphics{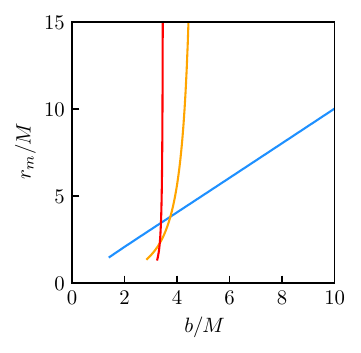}
\end{subfigure}
\begin{subfigure}[c]{0.32\textwidth} % Middle
\centering
\caption{$\Xi/M = 0$}
\includegraphics{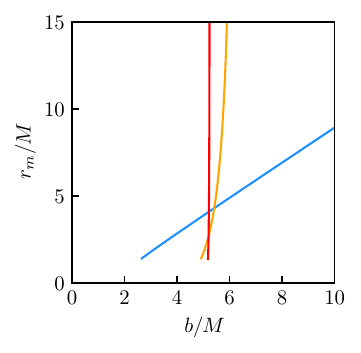}
\end{subfigure}
\begin{subfigure}[c]{0.32\textwidth} % Right
\centering
\caption{$\Xi/M = 1$}
\includegraphics{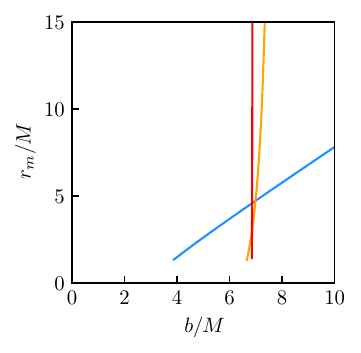}
\end{subfigure}

% SV wormhole
\begin{subfigure}[c]{0.32\textwidth} % Left
\centering
\includegraphics{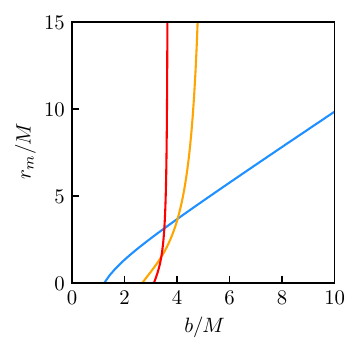}
\end{subfigure}
\begin{subfigure}[c]{0.32\textwidth} % Middle
\centering
\includegraphics{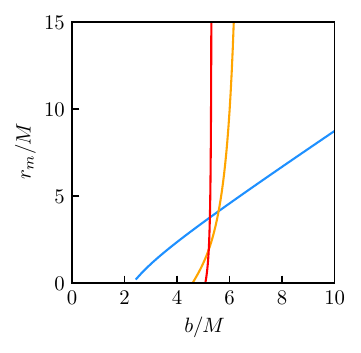}
\end{subfigure}
\begin{subfigure}[c]{0.32\textwidth} % Right
\centering
\includegraphics{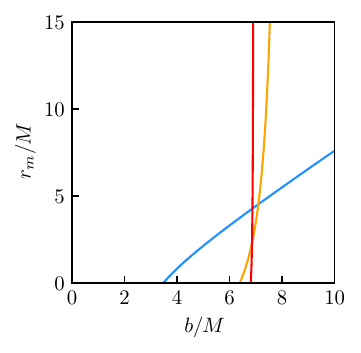}
\end{subfigure}
 
\caption{
First three transfer functions of a SV black hole with $\rt=1.5 M$ (top row) and a traversable wormhole with $\rt=2.5 M$ (bottom row) for different values of $\Xi$ per column.
We have used different colours for the direct (blue), lensed (orange) and photon ring (red) contributions.
Larger $\rt$ values widen the transfer functions of higher-order images, although they do not change the location of the critical impact parameter $b_c$.
On the other hand, as $\Xi$ grows, we observe that transfer functions shift to the right in addition to becoming steeper for the lensed and photon ring ones.
This results in an increase in $b_c$ for greater $\Xi$, as already predicted by Eq.~\eqref{eq:b_crit}.
}
\label{fig:SV_BH_transfer_functions}
\end{figure}
%%%%%%%%%%%%%%%%%%%%%%%%%%%%

We will assume that there are no light rays coming from the other side of the throat ($r<0$) so that the emission is only due to the accretion disk in our region with $r>0$.
Consequently, back-traced light rays falling into the throat of the wormhole will not make any contribution to the observed intensity, resulting in a central dark spot in the final image.

\begin{figure}[ht!]
\centering 
% SV black hole
\begin{subfigure}[c]{0.32\textwidth}
\caption{$\Xi/M = -1$}
\includegraphics[width=\textwidth]{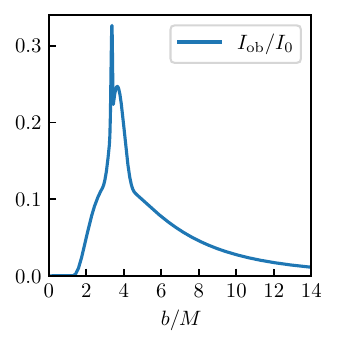}
\end{subfigure}
\begin{subfigure}[c]{0.32\textwidth}
\caption{$\Xi/M = 0$}
\includegraphics[width=\textwidth]{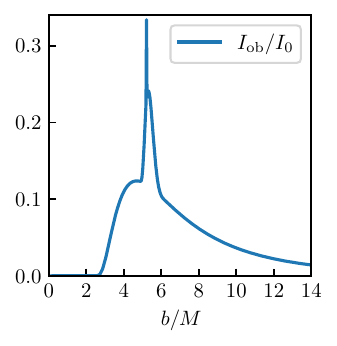}
\end{subfigure}
\begin{subfigure}[c]{0.32\textwidth}
\caption{$\Xi/M = 1$}
\includegraphics[width=\textwidth]{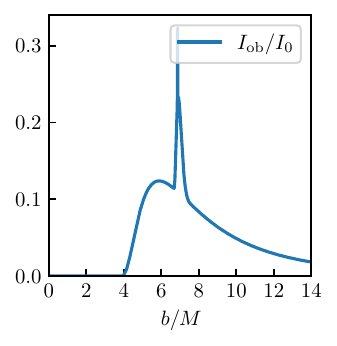}
\end{subfigure}

\begin{subfigure}[c]{0.32\textwidth}
\includegraphics[width=\textwidth]{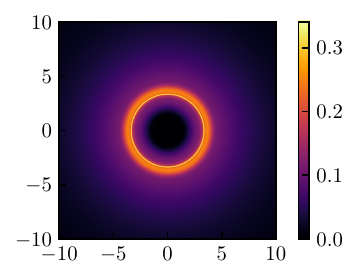}
\end{subfigure}
\begin{subfigure}[c]{0.32\textwidth}
\includegraphics[width=\textwidth]{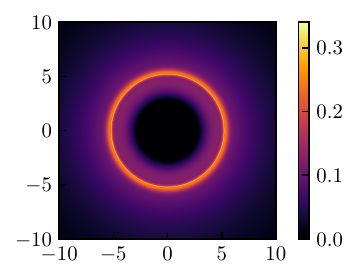}
\end{subfigure}
\begin{subfigure}[c]{0.32\textwidth}
\includegraphics[width=\textwidth]{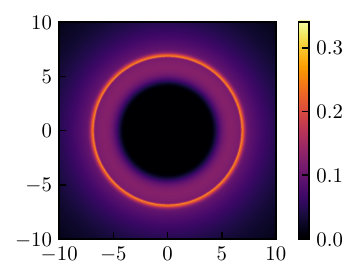}
\end{subfigure}

% SV wormhole
\begin{subfigure}[c]{0.32\textwidth}
\includegraphics[width=\textwidth]{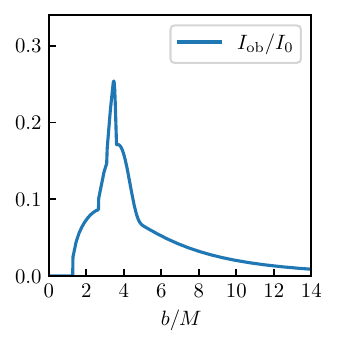}
\end{subfigure}
\begin{subfigure}[c]{0.32\textwidth}
\includegraphics[width=\textwidth]{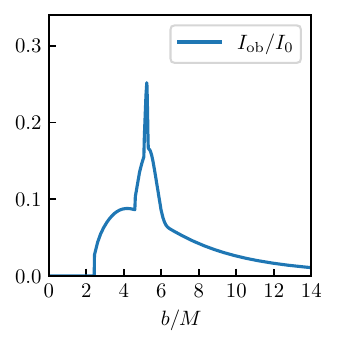}
\end{subfigure}
\begin{subfigure}[c]{0.32\textwidth}
\includegraphics[width=\textwidth]{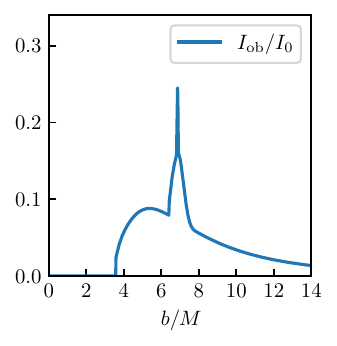}
\end{subfigure}

\begin{subfigure}[c]{0.32\textwidth}
\includegraphics[width=\textwidth]{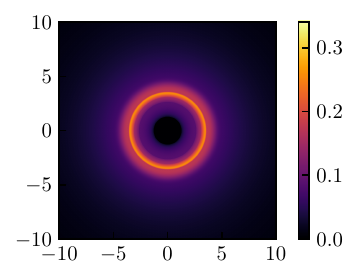}
\end{subfigure}
\begin{subfigure}[c]{0.32\textwidth}
\includegraphics[width=\textwidth]{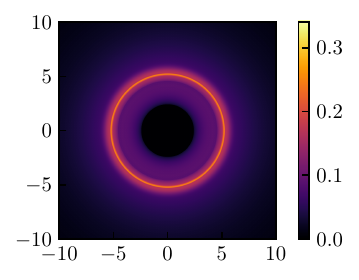}
\end{subfigure}
\begin{subfigure}[c]{0.32\textwidth}
\includegraphics[width=\textwidth]{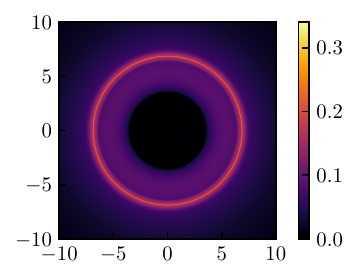}
\end{subfigure}

\caption{
Observed intensity profile and optical appearance of an SV black hole with $\rt = 1.5M$ (first two rows) and an SV traversable wormhole with $\rt = 2.5M$ (last two rows) for different values of $\Xi$ in each column.
The intensity profiles have been normalised to the maximum value $I_0$ of the emitted intensity.
Besides, all shadow plots use the same colour scale, which illustrates the demagnification of the lensing and photon rings as $\Xi$ increases.
}
\label{fig:SV_observed_intensities}
\end{figure}

Finally, we have calculated the observed intensities from the emission profiles in FIG.~\ref{fig:emission-models}, together with the optical appearances that these objects would cast in the sky from a face-on orientation. 
These results are shown in Fig~\ref{fig:SV_observed_intensities}, where we have considered six different cases for the dynamical Simpson-Visser geometry.
As already explained, these observed profiles are given by the sum of lensed and photon ring contributions overimposed on the direct emission coming from the thin disk.

Although black hole and wormhole images differ in the latter having wider light rings and a smaller shadow, the variation of $\Xi$ has similar effects on both of them.
The main observational consequence of the introduction of this parameter, which models an adiabatic temporal evolution, is an increase in the apparent size of the shadows as well as the diameter of observed light rings.
Additionally, we observe that these higher-order rings narrow as $\Xi$ grows.
The joint result of these two effects is a demagnification of lensing and photon ring contribution to the total observed intensity as $\Xi$ increases.
Despite this, lensed and photon rings are keep being noticeable in all figures.

%%%%%%%%% SHADOWS %%%%%%%%%
%%%%%%%%%%%%%%%%%%%%%%%%%%%

\subsection{Shadows of dynamical Hayward spacetime}\label{shadows-Hayward}

\begin{figure}[ht!]
\includegraphics[width=\textwidth]{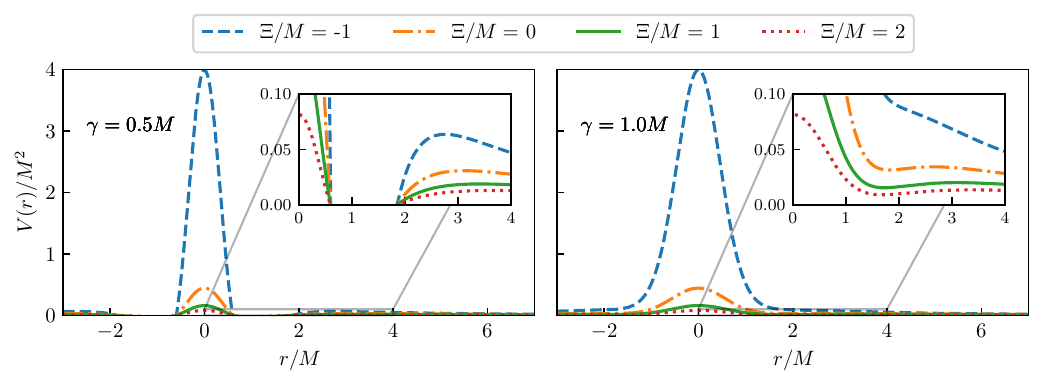}
\caption{
Hayward potential for $\gamma=0.5M$ (left) and $\gamma=1.0M$ (right) for different values of the parameter $\Xi$.
The parameter $\rt$ has been set to $\rt=1.5M$ in all cases.
The left panel corresponds to a two-horizon regular black hole, whereas the right panel corresponds to a traversable wormhole. 
In the former case, there are always two local maxima (photon spheres) for values of $r\geq0$, although the one at $r=0$ is hidden behind both horizons where $\e^{2\nu}$ vanishes. 
In the latter, there is a transition from having just one photon sphere at $r=0$ with $\Xi=-1$ to two photon spheres for the other values of $\Xi$. 
}
\label{fig:Hayward_potential}
\end{figure}

In this last example, we show the optical appearance of a dynamical Hayward spacetime, where $\e^{2\nu}$ is given by~\eqref{Hayward} and $\e^{2\xi}$ in Eq.~\eqref{eq:areal_case_1}.
The effective potential for null geodesics is then given by
\begin{align}
V(r) = \left( \sqrt{r^2+{\rt}^2}+\Xi \right)^{-2}\left( 1 - \frac{2Mr^2}{\left|r\right|^3 + 2M\gamma^2} \right)\, .
\end{align}
Since this potential depends on three parameters, we shall fix $\rt=1.5M$ so that there is always a throat, which will be hidden behind two event horizons $r_\pm$ whenever $\gamma<4\sqrt{3}M/9\approx 0.770M$.
Similarly to the Simpson-Visser example, we consider two values of $\gamma$ such that they result in two qualitatively different geometries: a two-horizon regular black hole $\gamma=0.5M$ and a horizonless wormhole $\gamma=1M$.
For each case, we give different values of $\Xi$ that model their optical appearances at different stages in their (adiabatic) evolution.
The shape of the potential for the considered cases is depicted in FIG.~\ref{fig:Hayward_potential}.

In the black hole case ($\gamma=0.5M$), the main difference from the Hayward potential is the presence of a global maximum at $r=0$, which means that we have a photon sphere at the throat in addition to the two unstable circular orbits on each side of the throat.
However, as this inner maximum is located at $r<r_+$ with $r_+$ in Eq.~\eqref{rpm},
it will not yield any additional rings in its optical appearance.
Therefore, light rays falling inside the horizon are trapped and do not return after orbiting that inner photon sphere, being then the outer light ring the only visible one, relative to the local maximum $\rp>r_+$.

\begin{figure}[ht!]
\centering
% Hayward black hole
\begin{subfigure}[c]{0.32\textwidth} % Left
\centering
\caption{$\Xi/M = -1$}
\includegraphics[width=\textwidth]{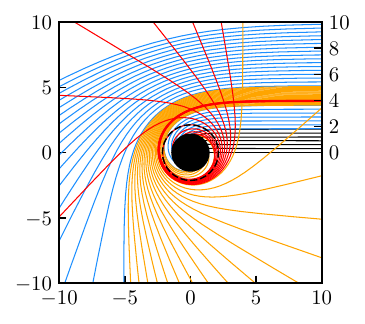}
\end{subfigure}
\begin{subfigure}[c]{0.32\textwidth} % Middle
\centering
\caption{$\Xi/M = 0$}
\includegraphics[width=\textwidth]{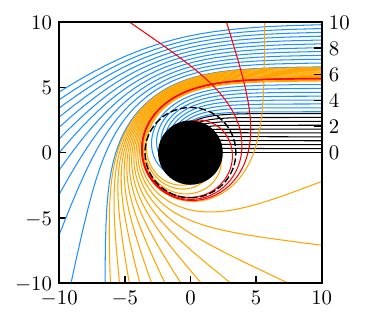}
\end{subfigure}
\begin{subfigure}[c]{0.32\textwidth} % Right
\centering
\caption{$\Xi/M = 1$}
\includegraphics[width=\textwidth]{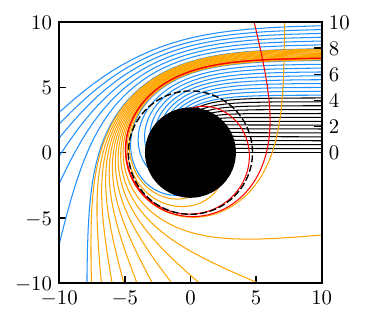}
\end{subfigure}

% Hayward wormhole
\begin{subfigure}[c]{0.32\textwidth} % Left
\centering
\includegraphics[width=\textwidth]{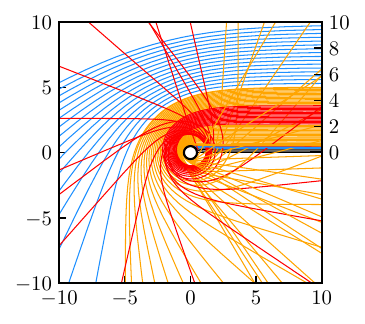}
\end{subfigure}
\begin{subfigure}[c]{0.32\textwidth} % Middle
\centering
\includegraphics[width=\textwidth]{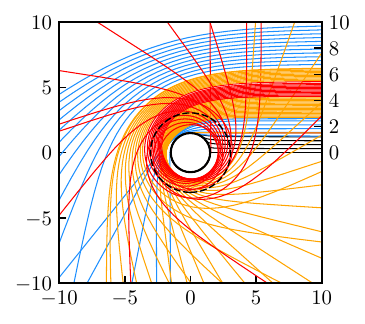}
\end{subfigure}
\begin{subfigure}[c]{0.32\textwidth} % Right
\centering
\includegraphics[width=\textwidth]{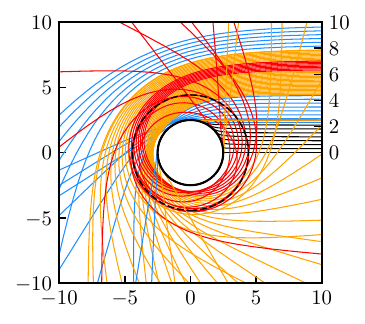}
\end{subfigure}
 
\caption{
Ray-tracing for the Hayward black hole with $\gamma=0.5M$ (top row) and traversable wormhole with $\gamma=1.0M$ (bottom row) for different values of $\Xi$ in each column ($\rt=1.5M$ in all cases).
They have been plotted in terms of the areal radius $\e^\xi$ and in units of $M=1$.
Light trajectories are coloured in blue, orange and red if they correspond to direct, lensed or photon ring images, respectively; while those coloured in black are those that do not hit the disk and make up the inner shadow.
In all cases, the dashed black circle represents the photon sphere $\rp$, while the central disk corresponds to the region inside the inner edge of the disk, filled with black if it is at the event horizon $r_\mathrm{hor}$ or with white if it is at the $\rt$.
Note that in the wormhole figures, geodesics cross each other due to the presence of two critical curves (middle and right), or one critical curve plus a \textit{strong lensing region} (left)
The effect of introducing the parameter $\Xi$ is the same as in the SV examples.
}
\label{fig:Hayward-ray-tracing}
\end{figure}

The values for the radial coordinate of the outer photon sphere $\rp$, as well as the critical impact parameter, must be found numerically, since the presence of non-vanishing parameters $\rt$ and $\Xi$ results in quite cumbersome expressions for the equations $V'(\rp)=0$ and $b_c=1/\sqrt{V(\rp)}$.
Nevertheless, as may be deduced from the potential FIG.~\ref{fig:Hayward_potential}, both $\rp$ and $b_c$ increase as $\Xi$ increases.
Furthermore, ray-tracing plots in FIG.~\ref{fig:Hayward-ray-tracing} show that both lensed and photon ring emission become narrower, resulting in a very similar behaviour to the Simpson-Visser black hole considered in Subsection~\ref{shadows-SV}.

The wormhole case ($\gamma=1M$) has some additional features that are not present in the Simpson-Visser wormhole and deserve further discussion.
Here, there are no horizons, and there is always an accessible photon sphere at the wormhole throat $\rp^1=0$.
Light rays near the critical curve with critical impact parameter $b_c^1 = |\rt+\Xi|$ approach this unstable circular orbit circle an arbitrarily large number of times before escaping back to infinity.
Those with $b\gtrsim b_c^1$ come back to our universe, whereas those with $b\lesssim b_c^1$ fall into the throat, circle arbitrarily many times and then go to $r\rightarrow-\infty$.
Additionally to this inner photon sphere, there may be an outer one $\rp^2$ at the local maximum, relative to that in the black hole case, whose values for $\rp^2$ and $b_c^2$ may only be obtained numerically (see \cite{Chiba:2017nml, Wei:2015qca} for a detailed discussion of this critical curve in the static case).
The main difference with the black hole case is that this outer photon sphere $\rp^2$ is not present for all values of $\Xi$, as the local maximum disappears for roughly $\Xi/M<-0.558$.
However, despite having only one critical curve at $b_c^1$, there is an additional \textit{strong lensing region} that results in an additional interval of impact parameters where light rays may also circle a large (but bounded) number of times, as can be seen in FIG.~\ref{fig:Hayward-ray-tracing}. 
It is also worth noting that, when both local maxima are present, there is always a local minimum or anti-photon sphere.

%%%%%%% RAY-TRACING %%%%%%%
%%%%%%%%%%%%%%%%%%%%%%%%%%%%

%%%% TRANSFER FUNCTIONS %%%%
\begin{figure}[ht!]
\centering
% Hayward black hole
\begin{subfigure}[c]{0.32\textwidth} % Left
\centering
\caption{$\Xi/M = -1$}
\includegraphics{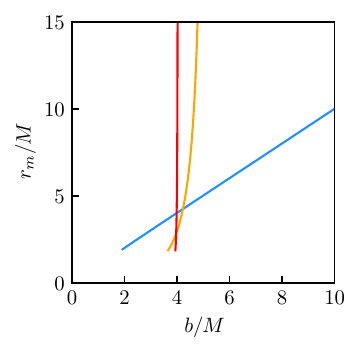}
\end{subfigure}
\begin{subfigure}[c]{0.32\textwidth} % Middle
\centering
\caption{$\Xi/M = 0$}
\includegraphics{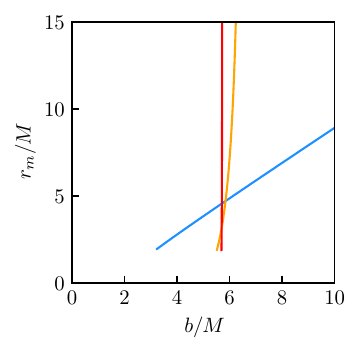}
\end{subfigure}
\begin{subfigure}[c]{0.32\textwidth} % Right
\centering
\caption{$\Xi/M = 1$}
\includegraphics{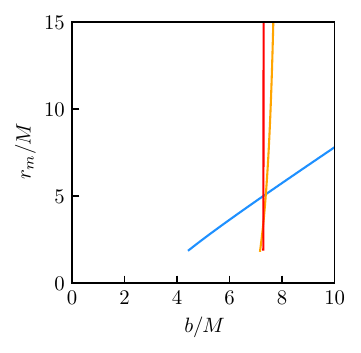}
\end{subfigure}

% Hayward wormhole
\begin{subfigure}[c]{0.32\textwidth} % Left
\centering
\includegraphics{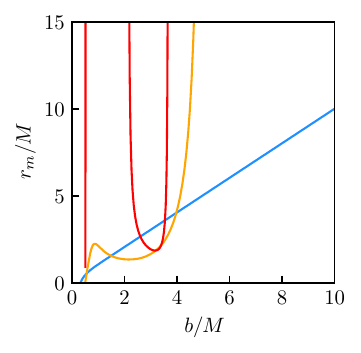}
\end{subfigure}
\begin{subfigure}[c]{0.32\textwidth} % Middle
\centering
\includegraphics{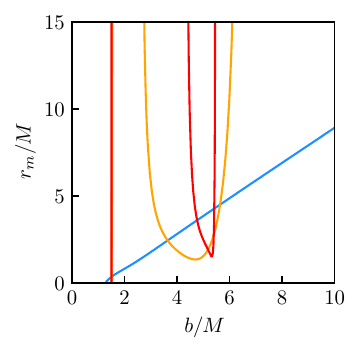}
\end{subfigure}
\begin{subfigure}[c]{0.32\textwidth} % Right
\centering
\includegraphics{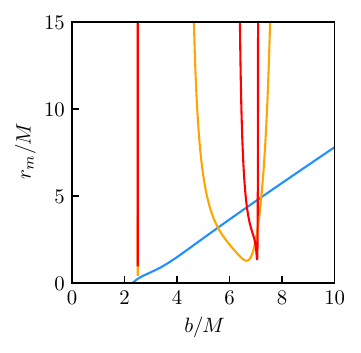}
\end{subfigure}
 
\caption{
First three transfer functions of a Hayward black hole with $\gamma=0.5 M$ (top row) and a traversable wormhole with $\gamma=1.0 M$ (bottom row) for different values of $\Xi$ per column ($\rt=1.5M$ in all cases).
We have used different colours for the direct (blue), lensed (orange) and photon ring (red) contributions.
In the middle and right figures in the bottom row, transfer functions for $m=2$ and $m=3$ have support on two disjoint regions due to the presence of an additional photon sphere.
As $\Xi$ grows, we observe that transfer functions shift to the right in addition to becoming steeper for lensed and photon ring contributions in the black hole case.
}
\label{fig:Hayward_BH_transfer_functions}
\end{figure}
%%%%%%%%%%%%%%%%%%%%%%%%%%%%

\begin{figure}[ht!]
\centering 
% SV black hole
\begin{subfigure}[c]{0.32\textwidth}
\caption{$\Xi/M = -1$}
\includegraphics[width=\textwidth]{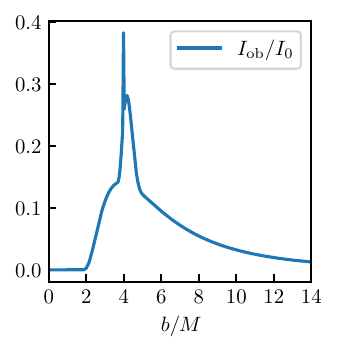}
\end{subfigure}
\begin{subfigure}[c]{0.32\textwidth}
\caption{$\Xi/M = 0$}
\includegraphics[width=\textwidth]{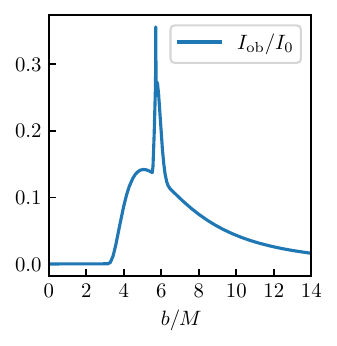}
\end{subfigure}
\begin{subfigure}[c]{0.32\textwidth}
\caption{$\Xi/M = 1$}
\includegraphics[width=\textwidth]{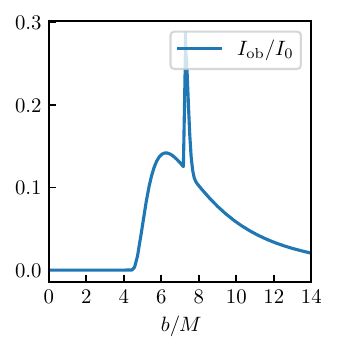}
\end{subfigure}

\begin{subfigure}[c]{0.32\textwidth}
\includegraphics[width=\textwidth]{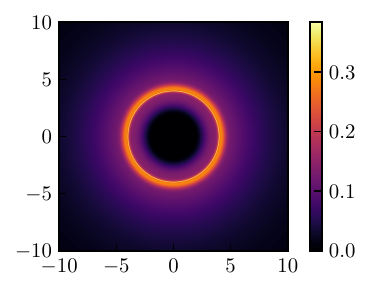}
\end{subfigure}
\begin{subfigure}[c]{0.32\textwidth}
\includegraphics[width=\textwidth]{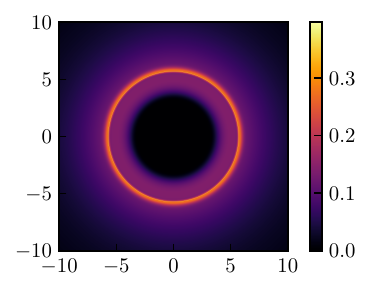}
\end{subfigure}
\begin{subfigure}[c]{0.32\textwidth}
\includegraphics[width=\textwidth]{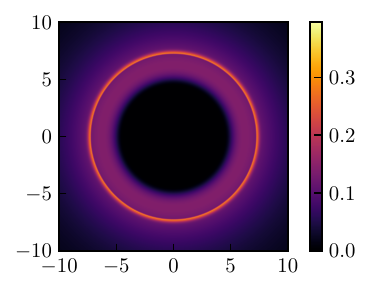}
\end{subfigure}

% SV wormhole
\begin{subfigure}[c]{0.32\textwidth}
\includegraphics[width=\textwidth]{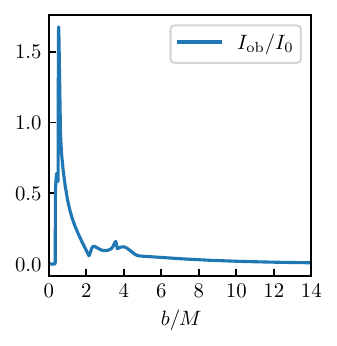}
\end{subfigure}
\begin{subfigure}[c]{0.32\textwidth}
\includegraphics[width=\textwidth]{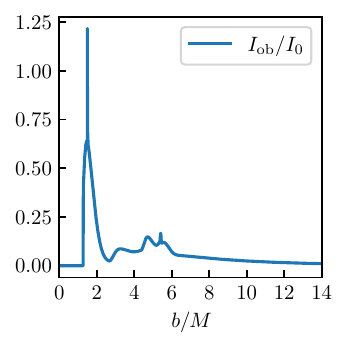}
\end{subfigure}
\begin{subfigure}[c]{0.32\textwidth}
\includegraphics[width=\textwidth]{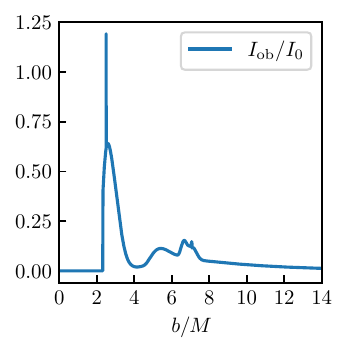}
\end{subfigure}

\begin{subfigure}[c]{0.32\textwidth}
\includegraphics[width=\textwidth]{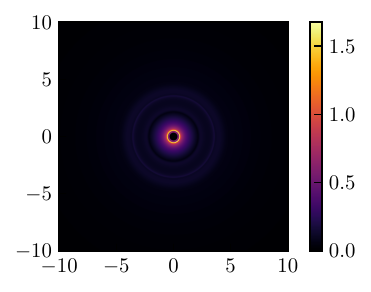}
\end{subfigure}
\begin{subfigure}[c]{0.32\textwidth}
\includegraphics[width=\textwidth]{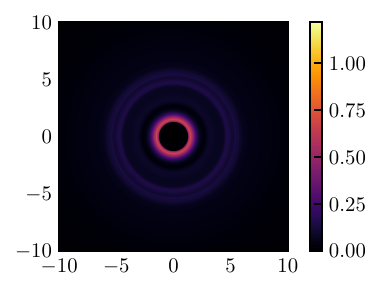}
\end{subfigure}
\begin{subfigure}[c]{0.32\textwidth}
\includegraphics[width=\textwidth]{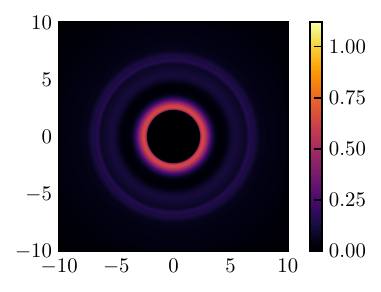}
\end{subfigure}

\caption{
Observed intensity profile and optical appearance of a Hayward regular black hole with $\gamma = 0.5M$ (first two rows) and a Hayward traversable wormhole with $\gamma = 1.0M$ (last two rows) for different values of $\Xi$ in each column.
The intensity profiles have been normalised to the maximum value $I_0$ of the emitted intensity.
}
\label{fig:Hayward_observed_intensities}
\end{figure}

In order to obtain the optical appearances, we use the emission profiles depicted in FIG.~\ref{fig:emission-models}.
The observed intensity is then computed for each of these examples using equation~\eqref{eq:observed-intensity} with transfer functions in FIG.~\ref{fig:Hayward_BH_transfer_functions}.
The observed intensities and optical appearances are shown in FIG.~\ref{fig:Hayward_observed_intensities}. 

In the Hayward black hole case, the optical appearances are qualitatively similar to those of the SV black hole: a single critical curve with lensing and photon rings becoming larger and narrower as $\Xi$ increases.
Additionally to these effects, the Hayward wormhole results in a very particular optical appearance, which differs in several aspects from the SV wormhole.

First, for $\Xi/M=0,1$, there are the usual lensed and photon rings superimposed on the direct emission due to the outer critical curve $b_c^2$.
Indeed, these rings appear even in the $\Xi/M=-1$ case, when that critical curve is no longer present.
As previously mentioned, the existence of a \textit{strong lensing} region allows light rays to intersect the accretion disk up to three times, thus producing these $m=2$ and $m=3$ contributions in the final image.

Furthermore, there is a wide intensity peak bordering the central brightness depression of the wormhole, which carries a large fraction of the total observed intensity.
This inner peak is part of the direct emission, and it is a key feature of horizonless geometries where the metric function $g_{tt}\rightarrow-1$ when $r\rightarrow0$, which causes the gravitational redshift to vanish and thus the observed intensity~\eqref{eq:observed-intensity} is not suppressed.
This fact, combined with an emission profile peaking near $r=0$, yields a very prominent emission which clearly distinguishes a Hayward wormhole from every other example considered here.

Lastly, there is a thin peak from the lensed emission superimposed on the above-mentioned peak, where the maximum of the observed intensity is found, and it is due to the inner critical curve at $b_c^1$.
However, this lensed ring is extremely narrow, carrying only a small fraction of the total intensity and being barely visible with the employed resolution\footnote{One may zoom in just to see a ring with a width of a few pixels.}.
In fact, the photon ring emission is so faint that it is completely negligible and cannot be perceived in the final output.
We remark that this Hayward wormhole has a very similar structure to the generalised black bounce introduced in \cite{Lobo:2020ffi}, whose optical appearance was studied in Refs.~\cite{Olmo:2021piq, Guerrero:2022qkh} is very similar to the one obtained here.

%%%%%%%%% SHADOWS %%%%%%%%%
%%%%%%%%%%%%%%%%%%%%%%%%%%%

\section{Conclusions} \label{conclusions}

In this paper, we have investigated time-dependent and spherically symmetric spacetimes. 
When we consider the black hole in a time-dependent or dynamical background, the horizon becomes a naked singularity for a large class of models. 
We have considered the prescription to avoid the singularity. 
An arbitrary dynamical and spherically symmetric spacetime can be realised by using Einstein's gravity coupled with two scalar fields \cite{Nojiri:2020blr}. 
Despite ghosts arise in the model given in Ref.~\cite{Nojiri:2020blr} in general, it has been shown that we can exclude the ghosts by using constraints~\cite{Nojiri:2023dvf, Nojiri:2023zlp, Elizalde:2023rds, Nojiri:2023ztz, Nojiri:2024dde, Alencar:2025nik}. 
By combining the formulation to construct a spherically symmetric spacetime and the prescription to avoid the singularity on the horizon, dynamical spacetimes where the wormhole is hidden inside the horizon(s) have been constructed analytically. 
The construction was also done for the dynamical extension of the Simpson-Visser black bounce~\cite{Simpson:2018tsi}.

The formulae for the separation of the energy-momentum tensor in the dynamical spacetime into the energy density and the pressures have been explicitly given, and by using the formulae, the energy conditions in the dynamical spacetime have been investigated. 
Although some conditions are not satisfied in the static spacetime, they can be satisfied in some dynamical cases. 
We should note that the violation of the energy conditions never means the instability of the system in the ghost-free two-scalar models. 

By investigating the thermodynamics of the Simpson-Visser black bounce, we have shown that although the thermodynamics is not changed from that for the Schwarzschild spacetime in the static case, 
even in the dynamical extension, the obtained thermodynamical quantities are completely identical to those of the static case. 
In \cite{Nojiri:2026ubx}, negative mass objects have also been constructed by using a two-scalar model and in \cite{Nojiri:2026auy}, the ghost-free Ellis-Bronnikov wormhole~\cite{Ellis:1973yv, Bronnikov:1973fh} was constructed as a candidate for negative mass objects and the optical appearance, like shadows, of such negative mass objects was compared with the Schwarzschild black hole and with the Simpson-Visser wormhole and it has been shown that there are clear differences in the photon ring substructure around the central object. 

Then 
in the last section, we have studied some observational features of these dynamical geometries.
In particular, we have simulated the optical appearances of non-stationary Simpson-Visser black bounce and Hayward spacetime.
We have assumed an adiabatic evolution, a simplification which allows us to integrate the null geodesic equation and the optical appearance is simulated via the ray-tracing method.
As a general result, we find that both the radial coordinates of unstable circular photon orbits $\rp$ and critical impact parameters $b_c$ increase with $\Xi_{02}(t)$, which in turn results in larger diameters for the lensed and photon rings, as well as a larger central brightness depression.
This fact, combined with a narrowing of these rings for larger values of $\Xi_{02}(t)$, has the final effect of demagnifying the lensing and photon ring contributions.

For the Simpson-Visser black bounce, we have considered two qualitatively different geometries: a black hole and a traversable wormhole.
Despite this, the final appearances are quite similar, with identical effects on the variation of $\Xi$.
Of course, this would no longer be the case if we had considered additional emission coming from the other side of the throat, which would produce extra rings in the wormhole images.

On the other hand, both Hayward examples, also a black hole and a wormhole, result in very distinct optical appearances.
While the Hayward black hole produces a shadow similar to those of the Simpson-Visser examples, the Hayward wormhole shows additional light rings due to an accessible maximum of the potential at $\rp^1=0$.
Even when this inner photon sphere is the only one present, exterior lensed and photon rings appear because of a \textit{strong lensing region}.
Theoretically, the non-existence of an outer critical curve would result in a finite set of rings instead of the usual infinite sequence of self-similar rings around the critical curve, as the number of disk intersections cannot be arbitrarily large.
However, from an observational point of view, as light rays in a \textit{strong lensing region} may hit the accretion disk several times, there are additional lensed and photon rings separated from those caused by the inner photon sphere.
This effect, together with the fact that higher-order rings contribute negligibly to the final image, makes it difficult to discriminate in practice whether a Hayward wormhole has two photon spheres or just the inner one plus a \textit{strong lensing region}.

\section*{Acknowledgements}

This work is supported by the Spanish National Grant PID2024-157196NB-I00 funded by MICIU/AEI/10.13039/501100011033 and by the Department of Education, Junta de Castilla y Le\'on and FEDER Funds, Ref. CLU-2023-1-05.

\appendix

\section{Connections and Curvatures}

In this section, the calculations of the connections and the curvatures are given for the arguments, after that. 

In the dynamical spherically symmetric spacetime in \eqref{ssWHmtrc}, the only non-vanishing connections are 
\begin{align}
\label{connectionsWH}
&\Gamma^t_{tt}=\dot\nu \, , \quad \Gamma^r_{tt} 
= \e^{-2(\lambda - \nu)}\nu' \, ,
\quad \Gamma^t_{tr}=\Gamma^t_{rt}=\nu'\, , \quad 
\Gamma^t_{rr} = \e^{2\lambda - 2\nu}\dot\lambda \, , \quad 
\Gamma^r_{tr} = \Gamma^r_{rt} = \dot\lambda \, , \quad 
\Gamma^r_{rr}=\lambda'\, ,\nonumber \\
%&\nonumber \\
&\Gamma^i_{jk} = \bar{\Gamma} ^i_{jk}\, ,\quad 
\Gamma^t_{ij}= \e^{-2\nu+2\xi}\dot\xi \bar{g}_{ij} \, , \quad
\Gamma^i_{tj}=\Gamma^i_{jt}= \dot\xi \delta^i_{\ j}\, , \quad 
\Gamma^r_{ij}=- \e^{-2\lambda+2\xi}\xi' \bar{g}_{ij} \, ,
\quad \Gamma^i_{rj}=\Gamma^i_{jr}= \xi' \delta^i_{\ j}\, ,
\end{align}
where $\bar{ \Gamma}^i_{jk}$ is the connection calculated from $\bar{g}_{ij}$, 
and an overdot and a prime denote differentiation with respect to $t$ and $r$, respectively. 
We use the following definition of the Riemann tensor 
\begin{align}
\label{Riemann}
R^\lambda_{\ \mu\rho\nu}
= -\Gamma^\lambda_{\mu\rho,\nu} + \Gamma^\lambda_{\mu\nu,\rho} - \Gamma^\eta_{\mu\rho}\Gamma^\lambda_{\nu\eta} + \Gamma^\eta_{\mu\nu}\Gamma^\lambda_{\rho\eta} \, ,
\end{align}
and we find 
\begin{align}
\label{curvaturesWH}
R_{rtrt} 
=&\, - \e^{2\lambda} \left\{ \ddot\lambda 
+ \left( \dot\lambda - \dot\nu \right) \dot\lambda \right\}
+ \e^{2\nu}\left\{ \nu'' + \left(\nu' - \lambda'\right)\nu' \right\} \, ,\nonumber \\
R_{titj} 
=&\, - \e^{2\xi} \bar{g}_{ij} \left\{ \ddot\xi + \left(\dot\xi - \dot\nu \right) \dot \xi - \e^{- 2\left(\lambda-\nu\right)} \nu' \xi' \right\} \, ,\nonumber \\
R_{rirj} 
=&\, \e^{2\xi} \bar{g}_{ij} \left\{ - \xi'' - \left( \xi' - \lambda' \right) \xi' + \e^{2\left(\lambda - \nu\right)} \dot\lambda \dot\xi \right\} \nonumber \\
R_{tirj} 
=&\, - \e^{2\xi} \bar{g}_{ij} \left\{ {\dot\xi}' + \dot\xi \xi' - \dot\xi \nu' - \xi' \dot\lambda \right\} \, , \nonumber \\
R_{ijkl} =&\, \e^{2\xi} \left(\bar{g}_{ik} \bar{g}_{jl} - \bar{g}_{il} \bar{g}_{jk} \right) \left\{ 2 + \e^{-2\nu+2\xi} {\dot\xi}^2 - \e^{-2\lambda+2\xi} {\xi'}^2 \right\} \, , \nonumber \\
R_{tt}=& - \left\{ \ddot\lambda + \left( \dot\lambda - \dot\nu \right) \dot\lambda \right\}
+ \e^{2\nu - 2\lambda} \left\{ \nu'' + \left(\nu' - \lambda'\right)\nu' \right\} 
- 2 \left\{ \ddot\xi + \left(\dot\xi - \dot\nu \right) \dot \xi - \e^{2\nu - 2\lambda} \nu' \xi' \right\} \, ,\nonumber \\
%&\nonumber \\ 
R_{rr} =& \, \e^{-2\left( \nu - \lambda \right)} \left\{ \ddot\lambda 
+ \left( \dot\lambda - \dot\nu \right) \dot\lambda \right\}
 - \left\{ \nu'' + \left(\nu' - \lambda'\right)\nu' \right\} 
+ 2 \left\{ - \xi'' - \left( \xi' - \lambda' \right) \xi' + \e^{2\left(\lambda - \nu\right)} \dot\lambda \dot\xi \right\} \, ,\nonumber \\ 
%&\nonumber \\ 
R_{tr} =& \, R_{rt} = - 2 \left\{ {\dot\xi}' + \dot\xi \xi' - \dot\xi \nu' - \xi' \dot\lambda \right\} \, , \nonumber \\
R_{ij} =&\, \left[ 2 + \e^{-2\nu+2\xi} {\dot\xi}^2 - \e^{-2\lambda+2\xi} {\xi'}^2 \right. \nonumber \\
&\, \left. + \e^{-2\nu + 2\xi} \left\{ \ddot\xi + \left(\dot\xi - \dot\nu \right) \dot \xi - \e^{2\nu- 2\lambda} \nu' \xi' \right\}
+ \e^{-2\lambda + 2\xi} \left\{ - \xi'' - \left( \xi' - \lambda' \right) \xi' + \e^{-2\nu + 2\lambda} \dot\lambda \dot\xi \right\} 
\right] \bar{g}_{ij}\, , \nonumber \\
%&\nonumber \\ 
R=& 2 \e^{-2 \nu} \left\{ \ddot\lambda 
+ \left( \dot\lambda - \dot\nu \right) \dot\lambda \right\} 
 - 2 \e^{-2\lambda}\left\{ \nu'' + \left(\nu' - \lambda'\right)\nu' \right\} \nonumber \\
&\, + 4 \e^{-2\nu} \left\{ \ddot\xi + \left(\dot\xi - \dot\nu \right) \dot \xi - \e^{- 2\left(\lambda-\nu\right)} \nu' \xi' \right\}
 - 4 \e^{-2\lambda} \left\{ \xi'' + \left( \xi' - \lambda' \right) \xi' - \e^{2\left(\lambda - \nu\right)} \dot\lambda \dot\xi \right\} \nonumber \\
&\, + 2 \e^{- 2\xi} \left\{ 2 + \e^{-2\nu+2\xi} {\dot\xi}^2 - \e^{-2\lambda+2\xi} {\xi'}^2 \right\} \, .
\end{align}
Here the Ricci tensor (Ricci curvature) $R_{\mu\nu}$ and the Ricci scalar (scalar curvature) $R$ are defined by $R_{\mu\nu} = R^\lambda_{\ \mu\lambda\nu} $ and $R=g^{\mu\nu}R_{\mu\nu}$. 
We use the above expressions to consider how consistently the dynamical solution appears. 

\bibliographystyle{apsrev4-1}\bibliography{References2}

@article{Nojiri:2020blr,
    author = "Nojiri, Shin'ichi and Odintsov, Sergei D. and Faraoni, Valerio",
    title = "{Searching for dynamical black holes in various theories of gravity}",
    eprint = "2010.11790",
    archivePrefix = "arXiv",
    primaryClass = "gr-qc",
    doi = "10.1103/PhysRevD.103.044055",
    journal = "Phys. Rev. D",
    volume = "103",
    number = "4",
    pages = "044055",
    year = "2021"
}

@article{Nojiri:2024dde,
    author = "Nojiri, Shin'ichi and Odintsov, S. D. and Folomeev, Vladimir",
    title = "{Wormholes inside stars and black holes}",
    eprint = "2401.15868",
    archivePrefix = "arXiv",
    primaryClass = "gr-qc",
    doi = "10.1103/PhysRevD.109.104007",
    journal = "Phys. Rev. D",
    volume = "109",
    number = "10",
    pages = "104007",
    year = "2024"
}

@article{Nojiri:2023dvf,
    author = "Nojiri, Shin'ichi and Nashed, G. G. L.",
    title = "{Wormhole solution free of ghosts in Einstein{\textquoteright}s gravity with two scalar fields}",
    eprint = "2309.12379",
    archivePrefix = "arXiv",
    primaryClass = "hep-th",
    doi = "10.1103/PhysRevD.108.124049",
    journal = "Phys. Rev. D",
    volume = "108",
    number = "12",
    pages = "124049",
    year = "2023"
}

@article{Nojiri:2023zlp,
    author = "Nojiri, Shin'ichi and Nashed, G. G. L.",
    title = "{Stable gravastar with large surface redshift in Einstein's gravity with two scalar fields}",
    eprint = "2310.16068",
    archivePrefix = "arXiv",
    primaryClass = "gr-qc",
    doi = "10.1088/1475-7516/2024/03/023",
    journal = "JCAP",
    volume = "03",
    pages = "023",
    year = "2024"
}

@article{Elizalde:2023rds,
    author = "Elizalde, E. and Nojiri, Shin'ichi and Odintsov, S. D. and Oikonomou, V. K.",
    title = "{Propagation of gravitational waves in a dynamical wormhole background for two-scalar Einstein{\textendash}Gauss{\textendash}Bonnet theory}",
    eprint = "2312.02889",
    archivePrefix = "arXiv",
    primaryClass = "gr-qc",
    doi = "10.1016/j.dark.2024.101536",
    journal = "Phys. Dark Univ.",
    volume = "45",
    pages = "101536",
    year = "2024"
}

@article{Nojiri:2023ztz,
    author = "Nojiri, Shin'ichi and Odintsov, Sergei D. and Sedrakian, Armen",
    title = "{Extremely small stars in scalar-tensor gravity: When stellar radius is less than Schwarzschild one}",
    eprint = "2312.15839",
    archivePrefix = "arXiv",
    primaryClass = "gr-qc",
    doi = "10.1016/j.nuclphysb.2024.116628",
    journal = "Nucl. Phys. B",
    volume = "1006",
    pages = "116628",
    year = "2024"
}

@article{Hayward:2005gi,
    author = "Hayward, Sean A.",
    title = "{Formation and evaporation of regular black holes}",
    eprint = "gr-qc/0506126",
    archivePrefix = "arXiv",
    doi = "10.1103/PhysRevLett.96.031103",
    journal = "Phys. Rev. Lett.",
    volume = "96",
    pages = "031103",
    year = "2006"
}

@article{Ellis:1973yv,
    author = "Ellis, H. G.",
    title = "{Ether flow through a drainhole - a particle model in general relativity}",
    doi = "10.1063/1.1666161",
    journal = "J. Math. Phys.",
    volume = "14",
    pages = "104--118",
    year = "1973"
}

@article{Bronnikov:1973fh,
    author = "Bronnikov, K. A.",
    title = "{Scalar-tensor theory and scalar charge}",
    journal = "Acta Phys. Polon. B",
    volume = "4",
    pages = "251--266",
    year = "1973"
}

@article{McVittie:1933zz,
    author = "McVittie, G. C.",
    title = "{The mass-particle in an expanding universe}",
    doi = "10.1093/mnras/93.5.325",
    journal = "Mon. Not. Roy. Astron. Soc.",
    volume = "93",
    pages = "325--339",
    year = "1933"
}

@article{Modesto:2025cre,
    author = "Modesto, Leonardo and Rattu, Edoardo",
    title = "{Do Black Holes Exist?}",
    eprint = "2510.04165",
    archivePrefix = "arXiv",
    primaryClass = "gr-qc",
    month = "10",
    year = "2025"
}

@article{Nashed:2024jqw,
    author = "Nashed, G. G. L. and Nojiri, Shin'ichi",
    title = "{General geometry realized by four-scalar model and application to f(Q) gravity}",
    eprint = "2402.12860",
    archivePrefix = "arXiv",
    primaryClass = "gr-qc",
    doi = "10.1016/j.dark.2024.101655",
    journal = "Phys. Dark Univ.",
    volume = "46",
    pages = "101655",
    year = "2024"
}

@article{Katsuragawa:2024bwm,
    author = "Katsuragawa, Taishi and Nojiri, Shin'ichi and Odintsov, Sergei D.",
    title = "{Future singularity in an anisotropic universe}",
    eprint = "2406.18368",
    archivePrefix = "arXiv",
    primaryClass = "gr-qc",
    reportNumber = "KEK-TH-2632, KEK-Cosmo-0348",
    doi = "10.1103/PhysRevD.110.064014",
    journal = "Phys. Rev. D",
    volume = "110",
    number = "6",
    pages = "064014",
    year = "2024"
}

@article{Katsuragawa:2025zcy,
    author = "Katsuragawa, Taishi and Nojiri, Shin'ichi and Odintsov, Sergei D.",
    title = "{Wormhole spacetimes in an expanding universe: Energy conditions and future singularities}",
    eprint = "2511.03275",
    archivePrefix = "arXiv",
    primaryClass = "gr-qc",
    reportNumber = "KEK-TH-2774, KEK-Cosmo-0396",
    doi = "10.1016/j.dark.2026.102213",
    journal = "Phys. Dark Univ.",
    volume = "51",
    pages = "102213",
    year = "2026"
}

@article{Nojiri:2026ubx,
    author = "Nojiri, Shin'ichi and Odintsov, S. D.",
    title = "{May Negative Mass Objects exist in the sky?}",
    eprint = "2602.15058",
    archivePrefix = "arXiv",
    primaryClass = "gr-qc",
    reportNumber = "KEK-TH-2802, KEK-Cosmo-0409",
    month = "2",
    year = "2026"
}

@article{Dzhunushaliev:2011xx,
    author = "Dzhunushaliev, V. and Folomeev, V. and Kleihaus, B. and Kunz, J.",
    title = "{A Star Harbouring a Wormhole at its Core}",
    eprint = "1102.4454",
    archivePrefix = "arXiv",
    primaryClass = "astro-ph.GA",
    doi = "10.1088/1475-7516/2011/04/031",
    journal = "JCAP",
    volume = "04",
    pages = "031",
    year = "2011"
}

@article{Dzhunushaliev:2012ke,
    author = "Dzhunushaliev, Vladimir and Folomeev, Vladimir and Kleihaus, Burkhard and Kunz, Jutta",
    title = "{Mixed neutron star-plus-wormhole systems: Equilibrium configurations}",
    eprint = "1203.3615",
    archivePrefix = "arXiv",
    primaryClass = "gr-qc",
    doi = "10.1103/PhysRevD.85.124028",
    journal = "Phys. Rev. D",
    volume = "85",
    pages = "124028",
    year = "2012"
}

@article{Dzhunushaliev:2014mza,
    author = "Dzhunushaliev, Vladimir and Folomeev, Vladimir and Kleihaus, Burkhard and Kunz, Jutta",
    title = "{Hiding a neutron star inside a wormhole}",
    eprint = "1401.7093",
    archivePrefix = "arXiv",
    primaryClass = "gr-qc",
    doi = "10.1103/PhysRevD.89.084018",
    journal = "Phys. Rev. D",
    volume = "89",
    number = "8",
    pages = "084018",
    year = "2014"
}

@article{Aringazin:2014rva,
    author = "Aringazin, Ascar and Dzhunushaliev, Vladimir and Folomeev, Vladimir and Kleihaus, Burkhard and Kunz, Jutta",
    title = "{Magnetic fields in mixed neutron-star-plus-wormhole systems}",
    eprint = "1412.3194",
    archivePrefix = "arXiv",
    primaryClass = "gr-qc",
    doi = "10.1088/1475-7516/2015/04/005",
    journal = "JCAP",
    volume = "04",
    pages = "005",
    year = "2015"
}

@article{Dzhunushaliev:2015sla,
    author = "Dzhunushaliev, Vladimir and Folomeev, Vladimir and Urazalina, Ajnur",
    title = "{Star-plus-wormhole systems with two interacting scalar fields}",
    eprint = "1506.03897",
    archivePrefix = "arXiv",
    primaryClass = "gr-qc",
    doi = "10.1142/S0218271815500972",
    journal = "Int. J. Mod. Phys. D",
    volume = "24",
    number = "14",
    pages = "1550097",
    year = "2015"
}

@article{Dzhunushaliev:2016ylj,
    author = "Dzhunushaliev, Vladimir and Folomeev, Vladimir and Kleihaus, Burkhard and Kunz, Jutta",
    title = "{Can mixed star-plus-wormhole systems mimic black holes?}",
    eprint = "1601.04124",
    archivePrefix = "arXiv",
    primaryClass = "gr-qc",
    doi = "10.1088/1475-7516/2016/08/030",
    journal = "JCAP",
    volume = "08",
    pages = "030",
    year = "2016"
}

@article{Dzhunushaliev:2022elv,
    author = "Dzhunushaliev, Vladimir and Folomeev, Vladimir and Kleihaus, Burkhard and Kunz, Jutta",
    title = "{Mixed neutron-star-plus-wormhole systems: Rotating configurations}",
    eprint = "2210.04425",
    archivePrefix = "arXiv",
    primaryClass = "gr-qc",
    doi = "10.1103/PhysRevD.107.044060",
    journal = "Phys. Rev. D",
    volume = "107",
    number = "4",
    pages = "044060",
    year = "2023"
}

@article{Simpson:2018tsi,
    author = "Simpson, Alex and Visser, Matt",
    title = "{Black-bounce to traversable wormhole}",
    eprint = "1812.07114",
    archivePrefix = "arXiv",
    primaryClass = "gr-qc",
    doi = "10.1088/1475-7516/2019/02/042",
    journal = "JCAP",
    volume = "02",
    pages = "042",
    year = "2019"
}

@article{Mazur:2001fv,
    author = "Mazur, Pawel O. and Mottola, Emil",
    title = "{Gravitational Condensate Stars: An Alternative to Black Holes}",
    eprint = "gr-qc/0109035",
    archivePrefix = "arXiv",
    reportNumber = "LA-UR-01-5067, LA-UR-01-5067",
    doi = "10.3390/universe9020088",
    journal = "Universe",
    volume = "9",
    number = "2",
    pages = "88",
    year = "2023"
}

@article{Dzhunushaliev:2013lna,
    author = "Dzhunushaliev, Vladimir and Folomeev, Vladimir and Kleihaus, Burkhard and Kunz, Jutta",
    title = "{Mixed neutron-star-plus-wormhole systems: Linear stability analysis}",
    eprint = "1302.5217",
    archivePrefix = "arXiv",
    primaryClass = "gr-qc",
    doi = "10.1103/PhysRevD.87.104036",
    journal = "Phys. Rev. D",
    volume = "87",
    number = "10",
    pages = "104036",
    year = "2013"
}

@article{Nariai:1999iok,
    author = "Nariai, Hidekazu",
    title = "{On a New Cosmological Solution of Einstein's Field Equations of Gravitation}",
    doi = "10.1023/A:1026602724948",
    journal = "Gen. Rel. Grav.",
    volume = "31",
    number = "6",
    pages = "963--971",
    year = "1999"
}

@article{Guerrero:2021ues,
    author = "Guerrero, Merce and Olmo, Gonzalo J. and Rubiera-Garcia, Diego and S{\'a}ez-Chill{\'o}n G{\'o}mez, Diego",
    title = "{Shadows and optical appearance of black bounces illuminated by a thin accretion disk}",
    eprint = "2105.15073",
    archivePrefix = "arXiv",
    primaryClass = "gr-qc",
    doi = "10.1088/1475-7516/2021/08/036",
    journal = "JCAP",
    volume = "08",
    pages = "036",
    year = "2021"
}

@article{Gralla:2019xty,
    author = "Gralla, Samuel E. and Holz, Daniel E. and Wald, Robert M.",
    title = "{Black Hole Shadows, Photon Rings, and Lensing Rings}",
    eprint = "1906.00873",
    archivePrefix = "arXiv",
    primaryClass = "astro-ph.HE",
    doi = "10.1103/PhysRevD.100.024018",
    journal = "Phys. Rev. D",
    volume = "100",
    number = "2",
    pages = "024018",
    year = "2019"
}

@article{Olmo:2021piq,
    author = "Olmo, Gonzalo J. and Rubiera-Garcia, Diego and  S{\'a}ez-Chill{\'o}n G{\'o}mez, Diego",
    title = "{New light rings from multiple critical curves as observational signatures of black hole mimickers}",
    eprint = "2110.10002",
    archivePrefix = "arXiv",
    primaryClass = "gr-qc",
    doi = "10.1016/j.physletb.2022.137045",
    journal = "Phys. Lett. B",
    volume = "829",
    pages = "137045",
    year = "2022"
}

@article{Perlick:2021aok,
    author = "Perlick, Volker and Tsupko, Oleg Yu.",
    title = "{Calculating black hole shadows: Review of analytical studies}",
    eprint = "2105.07101",
    archivePrefix = "arXiv",
    primaryClass = "gr-qc",
    doi = "10.1016/j.physrep.2021.10.004",
    journal = "Phys. Rept.",
    volume = "947",
    pages = "1--39",
    year = "2022"
}

@article{EventHorizonTelescope:2022wkp,
    author = "Akiyama, Kazunori and others",
    collaboration = "Event Horizon Telescope",
    title = "{First Sagittarius A* Event Horizon Telescope Results. I. The Shadow of the Supermassive Black Hole in the Center of the Milky Way}",
    eprint = "2311.08680",
    archivePrefix = "arXiv",
    primaryClass = "astro-ph.HE",
    doi = "10.3847/2041-8213/ac6674",
    journal = "Astrophys. J. Lett.",
    volume = "930",
    number = "2",
    pages = "L12",
    year = "2022"
}

@article{Paugnat:2022qzy,
    author = "Paugnat, Hadrien and Lupsasca, Alexandru and Vincent, Fr{\'e}d{\'e}ric and Wielgus, Maciek",
    title = "{Photon ring test of the Kerr hypothesis: Variation in the ring shape}",
    eprint = "2206.02781",
    archivePrefix = "arXiv",
    primaryClass = "astro-ph.HE",
    doi = "10.1051/0004-6361/202244216",
    journal = "Astron. Astrophys.",
    volume = "668",
    pages = "A11",
    year = "2022"
}

@article{Bekenstein:1972tm,
    author = "Bekenstein, J. D.",
    title = "{Black holes and the second law}",
    doi = "10.1007/BF02757029",
    journal = "Lett. Nuovo Cim.",
    volume = "4",
    pages = "737--740",
    year = "1972"
}

@article{Bekenstein:1973ur,
    author = "Bekenstein, Jacob D.",
    title = "{Black holes and entropy}",
    doi = "10.1103/PhysRevD.7.2333",
    journal = "Phys. Rev. D",
    volume = "7",
    pages = "2333--2346",
    year = "1973"
}

@article{Hawking:1975vcx,
    author = "Hawking, S. W.",
    editor = "Gibbons, G. W. and Hawking, S. W.",
    title = "{Particle Creation by Black Holes}",
    doi = "10.1007/BF02345020",
    journal = "Commun. Math. Phys.",
    volume = "43",
    pages = "199--220",
    year = "1975",
    note = "[Erratum: Commun.Math.Phys. 46, 206 (1976)]"
}

@article{Guerrero:2022qkh,
    author = "Guerrero, Merce and Olmo, Gonzalo J. and Rubiera-Garcia, Diego and  S{\'a}ez-Chill{\'o}n G{\'o}mez, Diego",
    title = "{Light ring images of double photon spheres in black hole and wormhole spacetimes}",
    eprint = "2202.03809",
    archivePrefix = "arXiv",
    primaryClass = "gr-qc",
    doi = "10.1103/PhysRevD.105.084057",
    journal = "Phys. Rev. D",
    volume = "105",
    number = "8",
    pages = "084057",
    year = "2022"
}

@article{Gralla:2020srx,
    author = "Gralla, Samuel E. and Lupsasca, Alexandru and Marrone, Daniel P.",
    title = "{The shape of the black hole photon ring: A precise test of strong-field general relativity}",
    eprint = "2008.03879",
    archivePrefix = "arXiv",
    primaryClass = "gr-qc",
    doi = "10.1103/PhysRevD.102.124004",
    journal = "Phys. Rev. D",
    volume = "102",
    number = "12",
    pages = "124004",
    year = "2020"
}

@article{Cardenas-Avendano:2022csp,
    author = "C{\'a}rdenas-Avenda{\~n}o, Alejandro and Lupsasca, Alexandru and Zhu, Hengrui",
    title = "{Adaptive analytical ray tracing of black hole photon rings}",
    eprint = "2211.07469",
    archivePrefix = "arXiv",
    primaryClass = "gr-qc",
    doi = "10.1103/PhysRevD.107.043030",
    journal = "Phys. Rev. D",
    volume = "107",
    number = "4",
    pages = "043030",
    year = "2023"
}

@article{Bisnovatyi-Kogan:2022ujt,
    author = "Bisnovatyi-Kogan, Gennady S. and Tsupko, Oleg Yu.",
    title = "{Analytical study of higher-order ring images of the accretion disk around a black hole}",
    eprint = "2201.01716",
    archivePrefix = "arXiv",
    primaryClass = "gr-qc",
    doi = "10.1103/PhysRevD.105.064040",
    journal = "Phys. Rev. D",
    volume = "105",
    number = "6",
    pages = "064040",
    year = "2022"
}

@article{Chiba:2017nml,
    author = "Chiba, Takeshi and Kimura, Masashi",
    title = "{A note on geodesics in the Hayward metric}",
    eprint = "1701.04910",
    archivePrefix = "arXiv",
    primaryClass = "gr-qc",
    doi = "10.1093/ptep/ptx037",
    journal = "PTEP",
    volume = "2017",
    number = "4",
    pages = "043E01",
    year = "2017"
}

@article{Wei:2015qca,
    author = "Wei, Shao Wen and Liu, Yu Xiao and Fu, Chun E.",
    title = "{Null Geodesics and Gravitational Lensing in a Nonsingular Spacetime}",
    eprint = "1510.02560",
    archivePrefix = "arXiv",
    primaryClass = "gr-qc",
    doi = "10.1155/2015/454217",
    journal = "Adv. High Energy Phys.",
    volume = "2015",
    pages = "454217",
    year = "2015"
}

@article{Lobo:2020ffi,
    author = "Lobo, Francisco S. N. and Rodrigues, Manuel E. and de Sousa Silva, Marcos V. and Simpson, Alex and Visser, Matt",
    title = "{Novel black-bounce spacetimes: wormholes, regularity, energy conditions, and causal structure}",
    eprint = "2009.12057",
    archivePrefix = "arXiv",
    primaryClass = "gr-qc",
    doi = "10.1103/PhysRevD.103.084052",
    journal = "Phys. Rev. D",
    volume = "103",
    number = "8",
    pages = "084052",
    year = "2021"
}

@article{Alencar:2025nik,
    author = "Alencar, G. and D{\'a}rlla, R. and Nojiri, Shin'ichi and Odintsov, Sergei D. and S{\'a}ez-Chill{\'o}n G{\'o}mez, Diego",
    title = "{Non-stationary wormholes with the presence of scalar fields and modified gravity}",
    eprint = "2508.05536",
    archivePrefix = "arXiv",
    primaryClass = "gr-qc",
    reportNumber = "KEK-TH-2742, KEK-Cosmo-0387",
    month = "8",
    year = "2025"
}

@article{Bronzwaer:2021lzo,
    author = "Bronzwaer, Thomas and Falcke, Heino",
    title = "{The Nature of Black Hole Shadows}",
    eprint = "2108.03966",
    archivePrefix = "arXiv",
    primaryClass = "astro-ph.HE",
    doi = "10.3847/1538-4357/ac1738",
    journal = "Astrophys. J.",
    volume = "920",
    number = "2",
    pages = "155",
    year = "2021"
}

@article{EventHorizonTelescope:2019dse,
    author = "Akiyama, Kazunori and others",
    collaboration = "Event Horizon Telescope",
    title = "{First M87 Event Horizon Telescope Results. I. The Shadow of the Supermassive Black Hole}",
    eprint = "1906.11238",
    archivePrefix = "arXiv",
    primaryClass = "astro-ph.GA",
    doi = "10.3847/2041-8213/ab0ec7",
    journal = "Astrophys. J. Lett.",
    volume = "875",
    pages = "L1",
    year = "2019"
}

@article{EventHorizonTelescope:2021dqv,
    author = "Kocherlakota, Prashant and others",
    collaboration = "Event Horizon Telescope",
    title = "{Constraints on black-hole charges with the 2017 EHT observations of M87*}",
    eprint = "2105.09343",
    archivePrefix = "arXiv",
    primaryClass = "gr-qc",
    reportNumber = "FERMILAB-PUB-21-847-PPD",
    doi = "10.1103/PhysRevD.103.104047",
    journal = "Phys. Rev. D",
    volume = "103",
    number = "10",
    pages = "104047",
    year = "2021"
}

@article{EventHorizonTelescope:2022xqj,
    author = "Akiyama, Kazunori and others",
    collaboration = "Event Horizon Telescope",
    title = "{First Sagittarius A* Event Horizon Telescope Results. VI. Testing the Black Hole Metric}",
    eprint = "2311.09484",
    archivePrefix = "arXiv",
    primaryClass = "astro-ph.HE",
    reportNumber = "FERMILAB-PUB-22-422-PPD",
    doi = "10.3847/2041-8213/ac6756",
    journal = "Astrophys. J. Lett.",
    volume = "930",
    number = "2",
    pages = "L17",
    year = "2022"
}

@article{Staelens:2023jgr,
    author = {Staelens, Seppe and Mayerson, Daniel R. and Bacchini, Fabio and Ripperda, Bart and K{\"u}chler, Lorenzo},
    title = "{Black hole photon rings beyond general relativity}",
    eprint = "2303.02111",
    archivePrefix = "arXiv",
    primaryClass = "gr-qc",
    doi = "10.1103/PhysRevD.107.124026",
    journal = "Phys. Rev. D",
    volume = "107",
    number = "12",
    pages = "124026",
    year = "2023"
}

@article{Kocherlakota:2023qgo,
    author = "Kocherlakota, Prashant and Rezzolla, Luciano and Roy, Rittick and Wielgus, Maciek",
    title = "{Prospects for future experimental tests of gravity with black hole imaging: Spherical symmetry}",
    eprint = "2307.16841",
    archivePrefix = "arXiv",
    primaryClass = "gr-qc",
    doi = "10.1103/PhysRevD.109.064064",
    journal = "Phys. Rev. D",
    volume = "109",
    number = "6",
    pages = "064064",
    year = "2024"
}

@article{Olmo:2023lil,
    author = "Olmo, Gonzalo J. and Rosa, Joao Luis and Rubiera-Garcia, Diego and Saez-Chillon Gomez, Diego",
    title = "{Shadows and photon rings of regular black holes and geonic horizonless compact objects}",
    eprint = "2302.12064",
    archivePrefix = "arXiv",
    primaryClass = "gr-qc",
    reportNumber = "IPARCOS-UCM-23-111",
    doi = "10.1088/1361-6382/aceacd",
    journal = "Class. Quant. Grav.",
    volume = "40",
    number = "17",
    pages = "174002",
    year = "2023"
}

@article{Olmo:2025ctf,
    author = "Olmo, Gonzalo J. and Rosa, Jo{\~a}o Lu{\'\i}s and Rubiera-Garcia, Diego and Rueda, Alejandro and S{\'a}ez-Chill{\'o}n G{\'o}mez, Diego",
    title = "{Shadows from thin accretion disks of parametrized black hole solutions}",
    eprint = "2507.16580",
    archivePrefix = "arXiv",
    primaryClass = "gr-qc",
    doi = "10.1103/s968-npmt",
    journal = "Phys. Rev. D",
    volume = "112",
    number = "8",
    pages = "084059",
    year = "2025"
}

@article{Nojiri:2026auy,
    author = "Nojiri, Shin'ichi and Odintsov, Sergei D. and S{\'a}ez-Chill{\'o}n G{\'o}mez, Diego",
    title = "{Observational signatures of negative mass wormholes through their shadows}",
    eprint = "2605.16177",
    archivePrefix = "arXiv",
    primaryClass = "gr-qc",
    reportNumber = "KEK-TH-2835, KEK-Cosmo-0421",
    doi = "10.1140/epjc/s10052-026-16172-3",
    journal = "Eur. Phys. J. C",
    volume = "86",
    number = "8",
    pages = "919",
    year = "2026"
}

\end{document}